\RequirePackage{ifpdf,xcolor,tikz}
\documentclass[hyper,letterpaper]{JHEP3}
\pdfoutput=1
\usepackage{amsmath,amssymb,amsfonts,bm,empheq}
\usepackage{cite}
\usepackage{subcaption}
\usepackage{nicefrac}
\usepackage[enableskew,vcentermath]{youngtab}
\usepackage{epstopdf}
\usepackage{url}
\usepackage{float}
\usepackage{slashed,framed,xcolor,tikz}
\newtheorem{proposition}{Proposition}[section]

\newtheorem{definition}[proposition]{Definition}

\def\printname#1{
        \if\draft y
                \smash{\makebox[0pt]{\hspace{-0.5in}
                        \raisebox{8pt}{\tt\tiny #1}}}
        \fi
}

\newlength{\standardunitlength}
\catcode`\@=11
\long\def\@makecaption#1#2{%
     \vskip 10pt

\setbox\@tempboxa\hbox{
       \small\sf{\bfcaptionfont #1. }\ignorespaces #2}%
     \ifdim \wd\@tempboxa >\captionwidth {%
         \rightskip=\@captionmargin\leftskip=\@captionmargin
         \unhbox\@tempboxa\par}%
       \else
         \hbox to\hsize{\hfil\box\@tempboxa\hfil}%
     \fi}
\font\bfcaptionfont=cmssbx10 scaled \magstephalf
\newdimen\@captionmargin\@captionmargin=2\parindent
\newdimen\captionwidth\captionwidth=\hsize
\catcode`\@=12

\newdimen\tableauside\tableauside=1.0ex
\newdimen\tableaurule\tableaurule=0.4pt
\newdimen\tableaustep
\def\phantomhrule#1{\hbox{\vbox to0pt{\hrule height\tableaurule width#1\vss}}}
\def\phantomvrule#1{\vbox{\hbox to0pt{\vrule width\tableaurule height#1\hss}}}
\def\sqr{\vbox{%
\phantomhrule\tableaustep
\hbox{\phantomvrule\tableaustep\kern\tableaustep\phantomvrule\tableaustep}%
\hbox{\vbox{\phantomhrule\tableauside}\kern-\tableaurule}}}
\def\squares#1{\hbox{\count0=#1\noindent\loop\sqr
\advance\count0 by-1 \ifnum\count0>0\repeat}}
\def\tableau#1{\vcenter{\offinterlineskip
\tableaustep=\tableauside\advance\tableaustep by-\tableaurule
\kern\normallineskip\hbox
    {\kern\normallineskip\vbox
      {\gettableau#1 0 }%
     \kern\normallineskip\kern\tableaurule}%
  \kern\normallineskip\kern\tableaurule}}
\def\gettableau#1 {\ifnum#1=0\let\next=\null\else
  \squares{#1}\let\next=\gettableau\fi\next}

\newcommand{\bea}{\begin{eqnarray}}
\newcommand{\eea}{\end{eqnarray}}
\newcommand{\be}{\begin{equation}}
\newcommand{\ee}{\end{equation}}

\newcommand{\ea}[1]{\begin{align}#1\end{align}}

\newcommand{\Z}{{\mathbb Z}}
\newcommand{\N}{{\mathbb N}}

\renewcommand{\bar}{\overline}

\newcommand\catalannumber[3]{
  \fill[white]  (#1) rectangle +(#2,#2);
  \fill[fill=gray!25]
  (#1)
  \foreach \dir in {#3}{
    \ifnum\dir=0
    -- ++(0,1)
    \else
    -- ++(1,0)
    \fi
  } |- (#1);
  \draw[help lines] (#1) grid +(#2,#2);
  \draw[dashed] (#1) -- +(#2,#2);
  \coordinate (prev) at (#1);
  \foreach \dir in {#3}{
    \ifnum\dir=0
    \coordinate (dep) at (0,1);
    \else
    \coordinate (dep) at (1,0);
    \fi
    \draw[line width=2pt,-stealth] (prev) -- ++(dep) coordinate (prev);
  };
}

\usetikzlibrary{calc}

\usepackage{mathtools}
\def\multiset#1#2{\ensuremath{\left(\kern-.1em\left(\genfrac{}{}{0pt}{}{#1}{#2}\right)\kern-.1em\right)}}

\title{Donaldson-Thomas invariants, torus knots,\\ and lattice paths}

\author{Mi{\l}osz Panfil$^{1}$, Marko Sto$\check{\text{s}}$i$\acute{\text{c}}$$^{2,3}$, and Piotr Su{\l}kowski$^{1,4}$
\\
$^1$ Faculty of Physics, University of Warsaw, ul. Pasteura 5, 02-093 Warsaw, Poland \\
$^2$ CAMGSD, Departamento de Matem\'atica, Instituto Superior T\'ecnico,
Av. Rovisco Pais, 1049-001 Lisboa, Portugal  \\
$^3$ Mathematical Institute SANU, Knez Mihailova 36, 11000 Beograd, Serbia \\
$^4$ Walter Burke Institute for Theoretical Physics, California Institute of Technology, Pasadena, CA 91125, USA 
}

\abstract{In this paper we find and explore the correspondence between quivers, torus knots, and combinatorics of counting paths. Our first result pertains to quiver representation theory -- we find explicit formulae for classical generating functions and Donaldson-Thomas invariants of an arbitrary symmetric quiver. We then focus on quivers corresponding to $(r,s)$ torus knots and show that their classical generating functions, in the extremal limit and framing $rs$, are generating functions of lattice paths under the line of the slope $r/s$. Generating functions of such paths satisfy extremal A-polynomial equations, which immediately follows after representing them in terms of the Duchon grammar. Moreover, these extremal A-polynomial equations encode Donaldson-Thomas invariants, which provides an interesting example of algebraicity of generating functions of these invariants. We also find a quantum generalization of these statements, i.e. a relation between motivic quiver generating functions, quantum extremal knot invariants, and $q$-weighted path counting. Finally, in the case of the unknot, we generalize this correspondence to the full HOMFLY-PT invariants and counting of Schr{\"o}der paths.
\\
\\
\\
\\
\\
\\
\\ 
\\
\\
\\
\\
{\tt CALT-2018-011}}

\begin{document}

\tableofcontents


\newpage

\section{Introduction}    \label{sec-intro}

Polynomial knot invariants, such as colored HOMFLY-PT polynomials, are quite involved functions of various variables. In this paper we  show that for a large class of $(r,s)$ torus knots, these polynomials admit a very simple combinatorial interpretation -- they are related to the counting of lattice paths under a line of a specific slope $r/s$. This immediately relates the field of knot theory to combinatorics and path counting problems. Furthermore, we relate this observation to the correspondence between knots and quivers, discovered recently in \cite{Kucharski:2017poe,Kucharski:2017ogk}, see also \cite{Kucharski:2016rlb,Luo:2016oza,Zhu:2017lsn,Stosic:2017wno}. One important consequence of the knots-quivers correspondence is an identification of Labastida-Mari{\~n}o-Ooguri-Vafa (LMOV) invariants \cite{OoguriV,Labastida:2000zp,Labastida:2000yw,Labastida:2001ts} with motivic Donaldson-Thomas invariants for quivers  \cite{Kontsevich:2010px,COM:8276935,Rei12,efimov2012}, which leads to the proof of integrality of a large class of LMOV invariants. Altogether these results lead to an intricate web of relations between knot invariants, combinatorics and path counting problems, string theory setup behind LMOV invariants, and representation theory of quivers.

Our first important result in this paper pertains to quiver representation theory; namely, in Proposition \ref{glavna} we provide an explicit formula for coefficients of a classical generating function associated to an arbitrary symmetric quiver. Such generating functions are of interest, because they encode numerical Donaldson-Thomas (DT) invariants. More precisely, numerical DT invariants can be extracted from the logarithm of such generating functions. As our second important result, in Proposition \ref{glavnalog} we provide a general formula for such a logarithm of the classical quiver generating series, which then leads to explicit formulae for numerical DT invariants of an arbitrary symmetric quiver. These results should be of interest to anyone interested in quiver representation theory and DT invariants, irrespective of all other relations to knots and counting paths that we discuss in this paper.

Having found general formulas for classical generating series and numerical DT invariants for arbitrary symmetric quivers, we then focus on quivers that via the knots-quivers correspondence are associated to $(r,s)$ torus knots in framing $rs$. In Proposition \ref{prop2} we show that classical generating functions for such quivers, which are equal to classical generating functions of colored extremal HOMFLY-PT polynomials of $(r,s)$ torus knots in framing $rs$, are also equal to generating functions of lattice paths under the line of the slope $r/s$.  Furthermore, in Proposition \ref{prop2-q} we find a quantum generalization of this statement, and relate to each other motivic generating functions of quivers, $q$-dependent generating functions of extremal colored HOMFLY-PT polynomials for torus knots, and $q$-weighted (by the area underneath) lattice paths. (Recall that extremal HOMFLY-PT polynomials are defined as coefficients of the highest or lowest powers of the variable $a$ of the full HOMFLY-PT polynomials \cite{Garoufalidis:2015ewa}.)

Analysis of generating functions of extremal colored knot polynomials brings into our game one other concept, namely that of (generalized, and extremal) A-polynomials. A-polynomials are algebraic curves associated to knots, and can be defined by certain algebraic equations, which are satisfied by classical generating functions of colored knot polynomials. Therefore, from the identification of knot polynomials and lattice path counting, it follows that generating functions of lattice paths should also satisfy A-polynomial equations (up to appropriate identification of parameters). We prove this statement by representing the path counting problem in terms of the Duchon grammar, and showing that it indeed leads to algebraic equations that agree with knot theoretic A-polynomials. From the viewpoint of LMOV and Donaldson-Thomas invariants, the fact that their generating functions satisfy algebraic equations is an example of algebraicity discussed in \cite{Mainiero:2016xaj}.

Subsequently, to illustrate the above claims, we find quivers that correspond to $(3,s)$ torus knots. From the knots-quivers correspondence we then know that these quivers encode  formulas for (extremal) colored HOMFLY-PT polynomials for $(3,s)$ torus knots; such explicit formulas have not been known before, therefore finding them is the next important result of this paper. Furthermore, it follows from Propositions \ref{prop2} and \ref{prop2-q} that these formulas also encode ($q$-weighted) generating functions of lattice paths under the lines of the slope $3/s$. Such formulas also have not been known before, so they provide yet another important result of this work.

Finally, we make the first step towards generalization of all these results from the extremal case to the full $a$-dependent HOMFLY-PT polynomials. We find such a generalization for the framed unknot, for which the lattice path counting turns out to be generalized to the counting of Schr{\"o}der paths. 

While the connection between torus knots, lattice paths, and quivers that we find is  new, it would interesting to understand if or how it relates to other combinatorial models of knot invariants, such as (Calabi-Yau) crystals discussed in \cite{HSS}, the representation of (uncolored) HOMFLY-PT polynomials in terms of motivic Donaldson-Thomas invariants discussed in \cite{Diaconescu-HOMFLY-DT}, the relations between path counting and uncolored bottom row HOMFLY-PT homology of torus knots \cite{gorsky}, yet another relation between Schr{\"o}der and superpolynomials discussed in \cite{DMMSS}, or combinatoral models for torus knots considered in \cite{ORSG,Bulycheva:2014nsa}.

The plan of this paper is as follows. In section \ref{sec-cast} we introduce relevant background: basics of knot invariants, the knots-quivers correspondence, and a summary of analytic combinatorics and lattice path counting. In section \ref{sec-DT} we find explicit formulae for classical generating functions and Donaldson-Thomas invariants for an arbitrary symmetric quiver. In section \ref{sec-knots-paths} we present the relation between invariants of torus knots and counting of lattice paths, and illustrate it from various perspective. In section \ref{sec-torus3p} we derive quivers and exact expressions for extremal colored HOMFLY-PT polynomials for a series of $(3,s)$ torus knots, which then lead to explicit expressions for the numbers of lattice paths under the lines of the slope $3/s$. Finally, in section \ref{sec-schroeder} we relate full $a$-dependent HOMFLY-PT polynomials of the unknot to the counting of Schr{\"o}der paths.


\section{Cast: knots, quivers, and paths}   \label{sec-cast}

In this section we present relevant background from three seemingly unrelated areas of research: knot invariants, quiver representation theory, and combinatorics of lattice paths. In the rest of the paper we will reveal surprising links between these topics.

\subsection{Knot invariants}

To start with we introduce relevant notation and briefly review those notions from knot theory, which will be of our main interest in the rest of the paper. We denote unreduced HOMFLY-PT polynomials as
\be
 \overline{P}_{R}(a,q) = \big\langle \textrm{Tr}_R U \big\rangle,
\ee
where the right hand side indicates that these polynomials arise as expectation values of Wilson loops in representation $R$ in Chern-Simons theory  \cite{Witten_Jones}, with $U=P\,\exp\oint_K A$ denoting the holonomy of $U(N)$ Chern-Simons gauge field along a knot $K$. This expectation value depends on the rank $N$ and the level of Chern-Simons theory, which are encoded in two parameters $a$ and $q$ of HOMFLY-PT polynomials. Unreduced polynomials are normalized so that
\be 
\overline{P}_{R}(a,q)=\overline{P}_R^{\bf 0_1} P_R(a,q),
\ee
where $P_R(a,q)$ is the corresponding reduced colored HOMFLY-PT polynomial (equal to 1 for the unknot), and $\overline{P}_R^{\bf 0_1}$ is the normalization factor of the unknot. 

Physical interpretation of knot polynomials in terms of Chern-Simons theory can be extended to topological string theory \cite{Witten:1992fb}. This interpretation led to an important Labastida-Mari{\~n}o-Ooguri-Vafa (LMOV) conjecture  \cite{OoguriV,Labastida:2000zp,Labastida:2000yw,Labastida:2001ts}, which states that colored HOMFLY-PT polynomials are encoded in certain integral invariants $N_{R,i,j}$, that in M-theory interpretation count bound states of M2 and M5-branes. These invariants are encoded in the Ooguri-Vafa operator
\be
Z(U,V) = \sum_R  \textrm{Tr}_R U \, \textrm{Tr}_R V = \exp\Big(  \sum_{n=1}^{\infty} \frac{1}{n} \textrm{Tr} U^n \textrm{Tr} V^n \Big),
\ee
where $V$ represents a source, and the sum runs over all two-dimensional partitions that label representations $R$. According to the LMOV conjecture, the expectation value of the Ooguri-Vafa operator provides a generating function of colored HOMFLY-PT polynomials and takes form
\be
\big\langle Z(U,V) \big\rangle = \sum_R \overline{P}_{R}(a,q) \textrm{Tr}_R V  = 
\exp \Big(  \sum_{n=1}^\infty \sum_R \frac{1}{n} f_{R}(a^n,q^n) \textrm{Tr}_R V^n  \Big).    \label{ZUV}
\ee
The functions $f_R(a,q)$ conjecturally encode integral invariants $N_{R,i,j}$ and take form
\be
f_{R}(a,q) = \sum_{i,j} \frac{N_{R,i,j} a^i q^j}{q-q^{-1}},  \label{fR}
\ee
and can be expressed as universal polynomials in colored HOMFLY-PT polynomials. Various tests of the LMOV conjecture have been conducted \cite{OoguriV,Labastida:2000zp,Labastida:2000yw,Ramadevi_Sarkar,Mironov:2017hde,Garoufalidis:2015ewa,Kucharski:2016rlb}, as well as an attempt of a proof \cite{Liu:2007kv}, but its general proof is still unknown. However, integrality of LMOV invariants for symmetric representations follows from the relation between knots and quivers and their relations to motivic Donaldson-Thomas invariants, as found recently in \cite{Kucharski:2017poe,Kucharski:2017ogk}.

Polynomial knot invariants have been generalized to the realm of knot homologies. First and important examples of such structures are Khovanov homology and Khovanov-Rozansky homology \cite{Khovanov,KhR1,KhR2}. It is believed that there exist knots homologies $\mathcal{H}^{S^r}_{i,j,k}$ for colored HOMFLY-PT polynomials, and various conjectural properties of those theories enable to determine corresponding colored superpolynomials for a large class of knots
\be
P_r (a,q,t) = \sum_{i,j,k} \, a^i q^j t^k \, \dim \mathcal{H}^{S^r}_{i,j,k}.    \label{superpolynomial}
\ee
For $t=-1$ these superpolynomials reduce to colored HOMFLY-PT polynomials. As we will summarize in what follows, knot homologies and superpolynomials play an important role in the relation to quivers too.

In this paper we are mainly interested in two simplifications of the above framework. First, we focus on symmetric representations $R=S^r$. This can be achieved by considering a one-dimensional source $V=x$, so that $\textrm{Tr}_R V \neq 0$ only for symmetric representations $R=S^r$, and then $\textrm{Tr}_{S^r}(x) = x^r$. Upon this specialization (\ref{ZUV}) reduces to the generating function of $S^r$-colored HOMFLY-PT polynomials $\overline{P}_{r}(a,q)\equiv \overline{P}_{S^r}(a,q)$ 
\be
P(x) = \langle Z(U,x) \rangle = \sum_{r=0}^\infty \overline{P}_{r}(a,q) x^r =   
\exp\Big({\sum_{r,n\geq 1} \frac{1}{n} f_{r}(a^n,q^n)x^{n r}}\Big),
\label{Pz2}
\ee
where $f_r(a,q)\equiv f_{S^r}(a,q)$ encode LMOV invariants denoted now $N_{r,i,j}\equiv N_{S^r,i,j}$,
\be
f_r(a,q) = \sum_{i,j} \frac{N_{r,i,j} a^i q^j}{q-q^{-1}}.
\ee
As mentioned above, these functions are universal polynomials in colored HOMFLY-PT polynomials, for example
\begin{align}
f_1(a,q) &= \overline{P}_1(a,q), \nonumber\\
f_2(a,q) &=  \overline{P}_2(a,q) - \frac{1}{2} \overline{P}_1(a,q)^2 -\frac{1}{2}  \overline{P}_1(a^2,q^2), \nonumber \\
f_3(a,q) &= \overline{P}_3(a,q) - \overline{P}_1(a,q)\overline{P}_2(a,q) + \frac{1}{3}\overline{P}_1(a,q)^3 - \frac{1}{3} \overline{P}_1(a^3,q^3), \nonumber
\end{align}
etc. The generating function (\ref{Pz2}) can be also rewritten in the product form
\be
P(x) = \prod_{r\geq 1;i,j;k\geq 0} \Big(1 - x^r a^i q^{j+2k+1} \Big)^{N_{r,i,j}}.
\label{Pr-LMOV} 
\ee
In the classical limit $q\to 1$ one can then define classical LMOV invariants\footnote{For fixed $r$ and $i$, the LMOV invariants $N_{r,i,j}$ are non-zero only for finitely many $j$ therefore making the sum in (\ref{bri}) finite.}
\be
n_{r,i}  = \sum_j N_{r,i,j},   \label{bri}
\ee
which are encoded in the following ratio
\be
y(x,a) =  \lim_{q\to 1} \frac{P(q^2x)}{P(x)} = 
\lim_{q\to 1}  \prod_{r\geq 1;i,j;k\geq 0} \Big(\frac{1 - x^r a^i q^{2(r+j+2k+1)} }{1 - x^r a^i q^{2(j+2k+1)}}\Big)^{N_{r,i,j}}  
= \prod_{r\geq 1;i} (1 - x^r a^i)^{-r\, n_{r,i}}.  \label{yxa}
\ee
Furthermore, $y=y(x,a)$ defined above satisfies an algebraic equation
\be
A(x,y)=0   \label{A-poly}
\ee
which is closely related to the augmentation polynomial, and it is also referred to as $a$-deformed A-polynomial \cite{AVqdef,Garoufalidis:2015ewa}. For $a=1$ it reduces to the original A-polynomial corresponding to a given knot.

The second simplification we consider amounts to taking the extremal limit \cite{Garoufalidis:2015ewa}. In this limit we focus on coefficients of extremal (highest or lowest) powers of variable $a$ of various knot invariants, such as colored HOMFLY-PT polynomials or superpolynomials. This limit is of particular interest for a large class of knots, whose colored HOMFLY-PT polynomials satisfy $\overline{P}_r(a,q) = \sum_{i=r \cdot c_-}^{r\cdot c_+} a^i p_{r,i}(q)$, for some fixed integers $c_\pm$ and for every natural number $r$, with $p_{r,r \cdot c_\pm}(q) \neq 0$. In this case, instead of the full colored HOMFLY-PT polynomial $P_r(a,q)$, we consider extremal polynomials, which depend then on a single variable $q$ and are denoted respectively $P^{\pm}_r(q)\equiv p_{r,r\cdot c_{\pm}}(q)$. We also introduce corresponding extremal LMOV invariants $N_{r,j}\equiv N_{r, r \cdot c_\pm,j}$ encoded in extremal functions $f^\pm_r(q)$, as well as associated classical extremal LMOV invariants $n^\pm_r$
\be
f^\pm_r(q) = \sum_j \frac{N_{r, r \cdot c_\pm,j} q^j}{q - q^{-1}},   \qquad\qquad
n^\pm_r = n_{r,r \cdot c_\pm} = \sum_j N_{r,r \cdot c_\pm,j}.         \label{br-minmax}
\ee
Extremal invariants $n^\pm_r$ satisfy improved integrality \cite{Garoufalidis:2015ewa}, i.e. they are divisible by $r$ -- this is an unexpected property, more general than M-theory integrality predictions. Furthermore, the generating series (\ref{Pz2}) in the extremal limit takes form 
\be
P^\pm(x) = \sum_{r=0}^\infty P^\pm_r(q) x^r 
= \prod_{r\geq 1;j;k\geq 0} \Big(1 - x^r q^{j+2k+1} \Big)^{N_{r,r \cdot c_\pm,j}},
\label{Pr-LMOV-minmax} 
\ee
while the ratio (\ref{yxa}) reduces to
\be
y(x) =  \lim_{q\to 1} \frac{P^{\pm}(q^2x)}{P^{\pm}(x)} = 
\lim_{q\to 1}  \prod_{r\geq 1;j;k\geq 0} \Big(\frac{1 - x^r  q^{(2r+j+2k+1)} }{1 - x^r q^{2(j+2k+1)}}\Big)^{N_{r, r \cdot c_\pm, j}}  
= \prod_{r\geq 1} (1 - x^r )^{-r n^\pm_{r}}.  \label{yxa-extremal}
\ee
If it is clear from the context which extremal invariants (minimal or maximal) we consider, we ignore the superscript $\pm$ and simply write $n_r\equiv n_r^{\pm}$. Extremal invariants $n_r$ can be extracted from the logarithmic derivative of $y(x)$. Indeed, if we denote
\be
x\frac{d}{dx}\log y(x) = x\frac{y'(x)}{y(x)} = \sum_{k=0}^{\infty} a_k x^k,   \label{log-y}
\ee
then
\be
n_r = \frac{1}{r^2} \sum_{d|r} \mu(d) a_{\frac{r}{d}},   \label{n-r}
\ee
where $\mu(d)$ is the M\"obius function. Note that integrality of $n_r$ implies that $\sum_{d|r} \mu(d) a_{\frac{r}{d}}$ is divisible by $r^2$, which is a nontrivial statement in number theory. Moreover, the function (\ref{yxa-extremal}) satisfies the extremal A-polynomial equation 
\be
A^\pm(x,y)=0,  \label{A-poly-extremal} 
\ee
whose coefficients are simply integer numbers (independent of $a$), and which can be found by appropriate rescaling of (\ref{A-poly}). Extremal A-polynomials have a number of interesting properties presented in \cite{Garoufalidis:2015ewa}, and in particular general formulas for extremal invariants $n_r$ can be deduced from the form of $A^{\pm}(x,y)$. Extremal A-polynomials also play a prominent role in this paper.


\subsection{Knots-quivers correspondence}   \label{ssec-knots-quivers}

\begin{figure}[t]
\begin{center}
\includegraphics[width=0.6\textwidth]{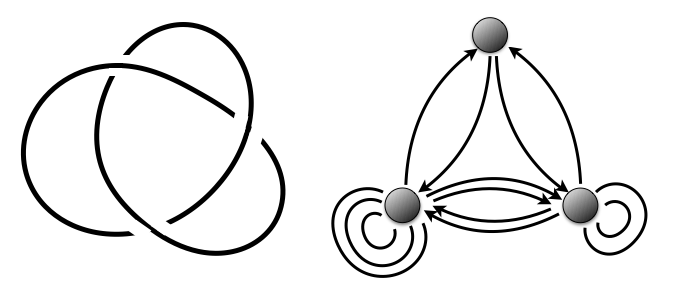} 
\caption{Trefoil knot and the corresponding quiver.}  \label{fig-trefoil}
\end{center}
\end{figure}

The correspondence between knots and quivers has been formulated in \cite{Kucharski:2017poe,Kucharski:2017ogk}. It states that to a given knot one can associate a symmetric quiver, in such a way, that various types of knot invariants are encoded in this corresponding quiver and in the moduli space of its representations. As an example, a quiver associated to trefoil knot is shown in fig. \ref{fig-trefoil}. Moduli spaces of quiver representations are characterized by various invariants, in particular numerical and motivic Donaldson-Thomas invariants \cite{Kontsevich:2010px,COM:8276935,Rei12}. In general such invariants are hard to compute, however they can be identified for some classes of quivers, in particular for symmetric quivers. Amusingly, these are symmetric quivers which play role in the knots-quivers correspondence. The knots-quivers correspondence was proven for all knots up to 6 crossings, infinite families of twist and torus knots, and some other examples in \cite{Kucharski:2017poe,Kucharski:2017ogk}, and for all rational knots in \cite{Stosic:2017wno}.

Consider a symmetric quiver with $m$ vertices. The structure of this quiver can be encoded in a symmetric square matrix $C\in\mathbb{Z}^{m\times m}$ with integer entries $C_{i,j}$, which denote the number of arrows from vertex $i$ to vertex $j$. The motivic generating series 
associated to this quiver is defined as
\begin{equation}
P_C(x_1,\ldots,x_m)=\sum_{d_1,\ldots,d_m} \frac{(-q)^{\sum_{i,j=1}^m C_{i,j}d_id_j}}{(q^2;q^2)_{d_1}\cdots(q^2;q^2)_{d_m}} x_1^{d_1}\cdots x_m^{d_m}.   \label{P-C}
\end{equation}
Motivic Donaldson-Thomas invariants $\Omega_{d_1,\ldots,d_m;j}$ of a symmetric quiver $Q$ can be interpreted as the intersection Betti numbers of the moduli space of all semisimple representations of $Q$, or as the Chow-Betti numbers of the moduli space of all simple representations \cite{MR,FR}, and they are encoded in the following product decomposition of the above series
\be
P_C(x_1,\ldots,x_m)=
\prod_{(d_1,\ldots,d_m)\neq 0} \prod_{j\in\mathbb{Z}} \prod_{k\geq 0} \Big(1 -  \big( x_1^{d_1}\cdots x_m^{d_m} \big) q^{j+2k+1} \Big)^{(-1)^{j+1}\Omega_{d_1,\ldots,d_m;j}}.   \label{PQx-Omega}
\ee
It is conjectured in \cite{Kontsevich:2010px} and proven in \cite{efimov2012} that $\Omega_{d_1,\ldots,d_m;j}$ are positive integers. 

One important manifestation of the knots-quivers correspondence is the statement, that generating functions of colored HOMFLY-PT polynomials (\ref{Pz2}) of a knot $K$ can be written in the form of the motivic generating function (\ref{P-C}) with some specific choice of a matrix $C$, and upon the identification
\be
x_i = x a^{a_i} q^{l_i} (-1)^{t_i+ C_{i,i}},     \label{specialize}
\ee
where $l_i=q_i-t_i$, and $a_i, q_i$ and $t_i$ are $(a,q,t)$--degrees of generators of the uncolored, reduced HOMFLY-PT homology of $K$. Therefore, it follows from the knots-quivers correspondence that the generating function of colored HOMFLY-PT polynomials can be written in the form
\be
P(x)  = \sum_{r=0}^{\infty} \overline{P}_r(a,q) x^r = \sum_{d_1,\ldots,d_m\geq 0} x^{d_1+\ldots + d_m}  q^{\sum_{i,j} C_{i,j} d_i d_j}    \frac{\prod_{i=1}^m q^{l_i d_i} a^{a_i d_i}(-1)^{t_i d_i}}{\prod_{i=1}^m(q^2;q^2)_{d_i}}.  \label{PxC}
\ee
Once general expressions for colored polynomials are known, after rewriting them in the above form, the matrix $C$ -- and thus the corresponding quiver -- can be identified. Moreover, the structure of the above formula is so constraining, that such a quiver can be identified even if only several colored polynomials are known. Note that it follows that all colored HOMFLY-PT polynomials for a given knot are encoded in a finite number of parameters: the matrix $C$ and parameters $a_i,q_i$ and $t_i$, which is a very strong prediction. Also recall, that from the quiver viewpoint a change of framing by $f$ simply amounts to adding $f$ to each element of the matrix $C$
\be
C \mapsto C+\left[ \begin{array}{ccc} 
f & f & \cdots \\ 
f & f & \cdots \\
\vdots & \vdots & \ddots
\end{array} \right]
\ee
It is also immediate to write down the generating series of extremal invariants (\ref{Pr-LMOV-minmax}) in the quiver form \cite{Kucharski:2017ogk}. It amounts to restricting a quiver to a subquiver, keeping only those vertices which are relevant in a given extremal limit. For such a smaller quiver $C$, with smaller number of vertices $m$, the change of variables (\ref{specialize}) simply does not involve $a$-dependnce
\be
x_i = x q^{l_i} (-1)^{t_i+C_{i,i}},     \label{specialize-ext}
\ee
and, analogously to (\ref{PxC}), in the extremal limit we get
\be
P^\pm(x) = \sum_{r=0}^\infty P^\pm_r(q) x^r =  \sum_{d_1,\ldots,d_m\geq 0} x^{d_1+\ldots + d_m}  q^{\sum_{i,j} C_{i,j} d_i d_j}    \frac{\prod_{i=1}^m q^{l_i d_i} 
(-1)^{t_i d_i}}{\prod_{i=1}^m(q^2;q^2)_{d_i}}.  \label{PxC-ext}
\ee

Furthermore, recall that in order to define classical LMOV invariants we considered the ratio of generating functions of colored HOMFLY-PT polynomials (\ref{yxa}), or (\ref{yxa-extremal}) in the extremal case. An analogous, albeit more general ratio can be considered for quiver generating functions
\begin{equation}
\frac{P_C(q^2 x_1,\ldots,q^2 x_m)}{P_C(x_1,\ldots,x_m)}=\sum_{l_1,\ldots,l_m} b_{l_1,\ldots,l_m}(q) x_1^{l_1}\ldots x_m^{l_m}.   \label{definv}
\end{equation}
Factorization of this ratio in the classical limit $q\to 1$ enables to define classical coefficients $b_{l_1,\ldots,l_m}\equiv b_{l_1,\ldots,l_m}(1)$ and numerical Donaldson-Thomas invariants $\Omega_{d_1, \dots, d_m}$ 
\be
y(x_1,\ldots,x_m) = \sum_{l_1,\ldots,l_m} b_{l_1,\ldots,l_m} x_1^{l_1}\ldots x_m^{l_m} = \prod_{(d_1,\ldots,d_m)\neq 0}  \big(1 -  x_1^{d_1}\cdots x_m^{d_m}  \big)^{\Omega_{d_1, \dots, d_m}}.    \label{definv-class}
\ee
Numerical Donaldson-Thomas invariants are combinations of their motivic counterparts
\be
\Omega_{d_1, \dots, d_m} = \left(d_1 + \dots + d_m \right)\sum_j (-1)^j\Omega_{d_1, \dots, d_m;j}.
\ee
We can also consider the specialization $x=x_1=\ldots = x_m$  and introduce diagonal sums $B_n$ of coefficients $b_{l_1,\ldots,l_m}$, in terms of which the generating function (\ref{definv-class}) reduces to
\be
y(x)\equiv y(x,\ldots,x) = \sum_{n=0}^{\infty} B_n x^n,  \qquad\qquad B_n = \sum_{l_1+\ldots + l_m=n} b_{l_1,\ldots,l_m}.    \label{yx-Bn}
\ee
Similarly, upon this specialization (and in analogy to (\ref{yxa-extremal})) we introduce diagonal DT invariants $n_r$
\be
y(x) \equiv y(x,\ldots,x) = \prod_{r=1}^{\infty}  \big(1 -  x^r  \big)^{-rn_r},  \qquad \qquad n_r = -\sum_{d_1+\ldots+ d_m = r}  \sum_j (-1)^j\Omega_{d_1, \dots, d_m;j}.  \label{nr-DT}
\ee

We also note that, while relating quiver generating functions to generating functions of colored knot polynomials, for a knot $K$ associated to a quiver $C$, it is natural to consider a modified quiver, encoded in a matrix $\bar{C}$ defined by
\be
\bar{C}_{i,j} = \left\{ \begin{array}{cl} 
-C_{i,j}+1 & \quad\textrm{for}\ i=j \\ 
-C_{i,j} &   \quad\textrm{for}\ i\neq j 
\end{array} \right.
\ee
Polynomials defined by the generating series of the form (\ref{PxC}), however with $C$ replaced by such a modified quiver $\bar{C}$ 
\begin{equation}
P'_{\bar{C}}(x_1,\ldots,x_m)=\sum_{d_1,\ldots,d_m} \frac{q^{\sum_{i,j=1}^m \bar{C}_{i,j}d_id_j}}{(q^2;q^2)_{d_1}\ldots(q^2;q^2)_{d_m}} x_1^{d_1}\ldots x_m^{d_m},
\end{equation}
are colored HOMFLY-PT polynomials of a knot $\bar{K}$, which is the mirror image of the original knot $K$. In this work we take advantage of the fact, that coefficients of the following quotient of generating series associated to $\bar{C}$
\begin{equation}
\frac{P'_{\bar{C}}(x_1,\ldots,x_m)}{P'_{\bar{C}}(q^2x_1,\ldots,q^2x_m)}=\sum_{l_1,\ldots,l_m} \bar{b}_{l_1,\ldots,l_m}(q) x_1^{l_1}\ldots x_m^{l_m},
\end{equation}
in the classical limit satisfy
\begin{equation}
{b}_{l_1,\ldots,l_m} =\bar{b}_{l_1,\ldots,l_m}.    \label{b-bbar}
\end{equation}
More generally, we postulate that the equality with the full $q$-dependence also holds
\be
{b}_{l_1,\ldots,l_m}(q) =\bar{b}_{l_1,\ldots,l_m}(q^{-1}).
\ee

As the framing plays a crucial role in this work, let us clarify in which choice we are primarily interested in. The quiver matrix for the bottom row of the right-handed (i.e. with all crossings positive) trefoil (i.e. $(2,3)$ torus) knot in framing 0 and framing $f=-6$ takes form respectively
\be
C^{(2,3)} = \left[
\begin{array}{cc}
2 & 1\\
1 & 0
\end{array}
\right], \qquad\qquad
C^{(2,3)}_{f=-6} = \left[
\begin{array}{cc}
-4 & -5\\
-5 & -6
\end{array}
\right].
\ee
Therefore the quiver matrix for the top row of the mirror (left-handed) trefoil, in framing 0 and framing $f=6$ reads respectively
\be
\overline{C}^{(2,3)} = \left[
\begin{array}{cc}
-1 & -1\\
-1 & 1
\end{array}
\right], \qquad\qquad
\overline{C}^{(2,3)}_{f=6} = \left[
\begin{array}{cc}
5 & 5\\
5 & 7
\end{array}
\right].
\ee
In the rest of the paper, unless otherwise stated, we consider top rows of left-handed torus knots, and denote their quiver matrices simply by $C$ (without bar). The framing $rs$ of the $(r,s)$ torus knot invoked in our main Proposition \ref{prop2} corresponds to this convention -- so in the above example this is $\overline{C}^{(2,3)}_{f=6}$ which makes contact with path counting (and in the rest of the paper we skip the bar on $C$). In view of (\ref{b-bbar}) the same results arise for $C^{(2,3)}_{f=-6}$, and in this convention the framing should be chosen as $-rs$. Moreover, we usually reorder entries of $C$ in such a way, that the top left element is the largest, see (\ref{C-22p1}) and (\ref{CT23}).


\subsection{Counting of lattice paths}    \label{ssec-paths}

\begin{figure}[h]
\centering
\includegraphics[scale=0.29]{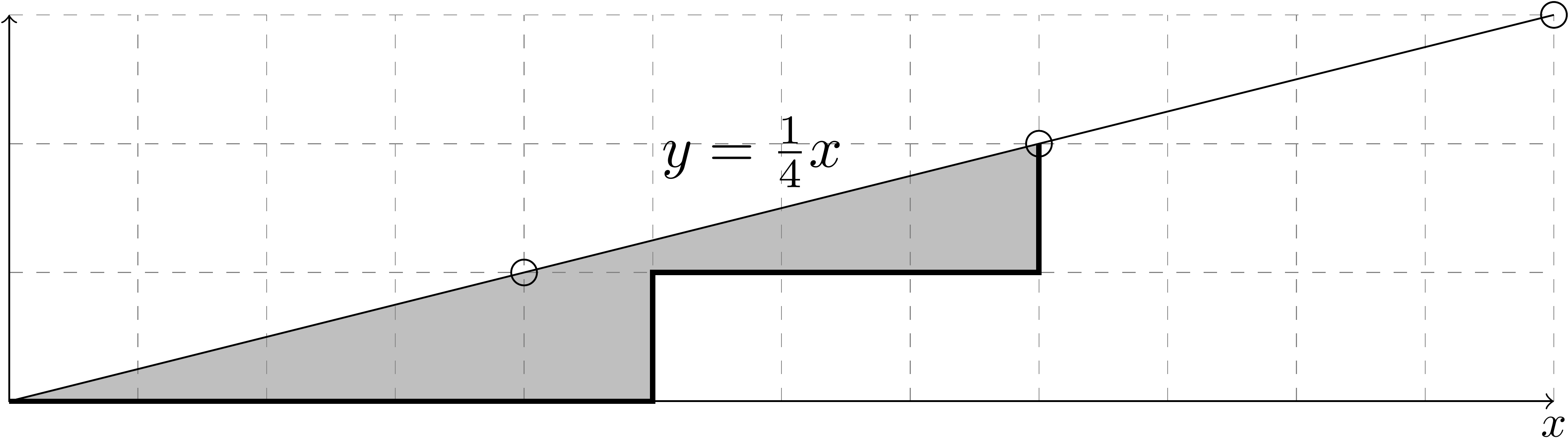}
 \caption{A lattice path under the line $y=\frac14 x$, and a shaded area between the path and the line.}  \label{14path}
\end{figure}

We discuss now the problem of counting of lattice paths. This is one of the basic problems in combinatorics, see e.g. \cite{banderier200237}. Consider a square lattice (with lattice spacing 1), and a line through the origin of a rational slope $r/s$, with mutually prime positive integers $r$ and $s$. This line passes through integer lattice points $(sk,rk)$ for all non-negative integers $k$. A basic question in combinatorics is how many different paths, made of elementary steps $(1,0)$ and $(0,1)$, one can draw between the origin and a given point $(sk,rk)$, in the wedge between the horizontal axis and the $y=\frac{r}{s}x$ line. It is then natural to consider the generating series that encodes the numbers of such paths for all $k$
\be
y_P(x) = \sum_{k=0}^{\infty} \sum_{\pi\in  \, k\textrm{-paths}} x^k = \sum_{k=0}^{\infty} c_k(1) x^k,  \label{Paths-x}
\ee
where $k$-paths in the second summation denotes the above mentioned paths from $(0,0)$ to $(sk,rk)$. It is also natural to consider a generalized counting, with each path $\pi$ weighted by the area $area(\pi)$ of the region between this path and the $x$-axis, and the corresponding $q$-deformed generating function
\be
y_{qP}(x) = \sum_{k=0}^{\infty} \sum_{\pi\in\,  k\textrm{-paths}} q^{area(\pi)}  x^k = \sum_{k=0}^{\infty} c_k(q) x^k.     \label{yqPx}
\ee
An example of a lattice path under the line $y=\frac{1}{4}x$, between points $(0,0)$ and $(8,2)$, is shown in Fig. \ref{14path}.

\begin{figure}[h]
\centering
\includegraphics[scale=0.27]{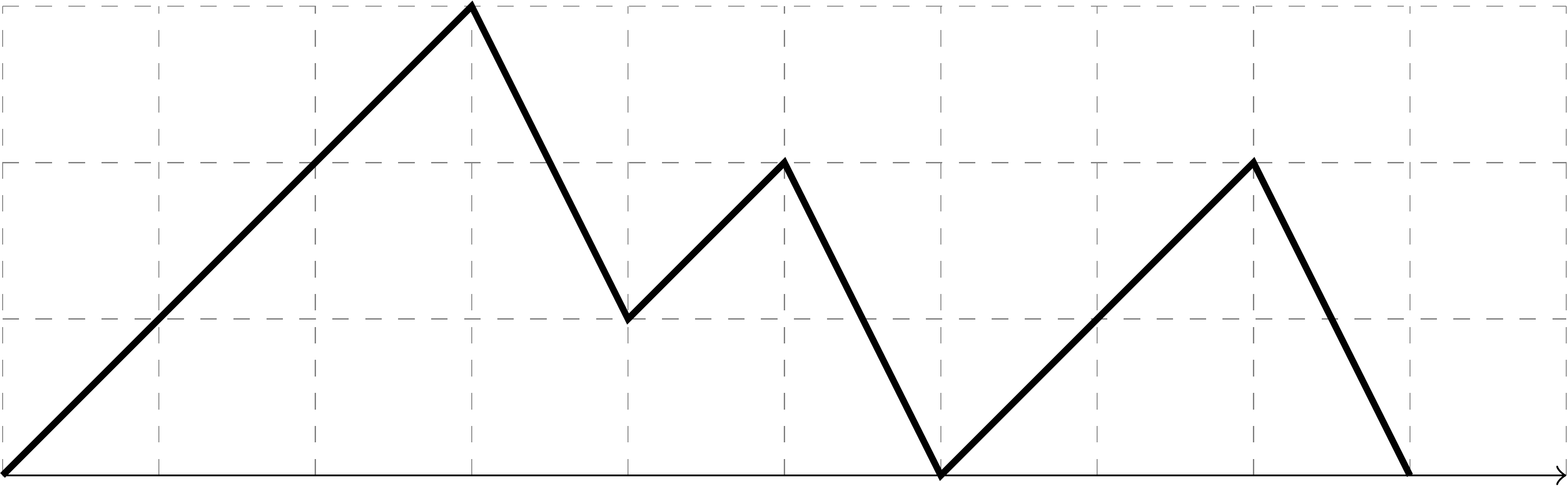}
 \caption{Counting of paths under the line $y=\frac12 x$ is equivalent to counting \emph{excursions}, i.e. paths in the upper half plane, made of elementary steps $(1,1)$ and $(1,-2)$.}  
\label{12path}
\end{figure}

\begin{figure}[h]
\centering
\includegraphics[scale=0.27]{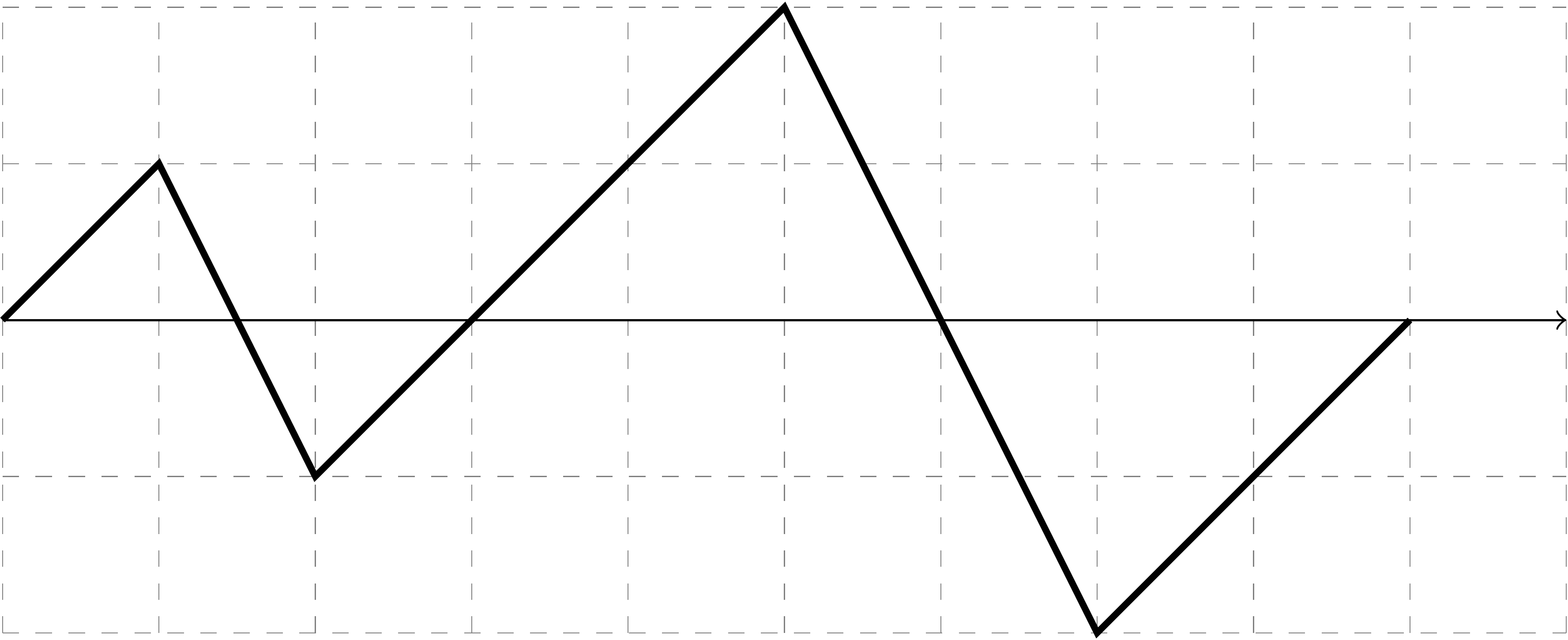}
\caption{An example of a \emph{bridge} (an unconstrained analog of an excursion) made of the same elementary steps as the excursion in fig. \protect\ref{12path}.}  
\label{12bridge}
\end{figure}

The above counting is equivalent to the counting of all paths in the upper half of the square lattice, starting at the origin and ending on the $y=0$ line, made of elementary steps $(1,r)$ and $(1,-s)$. For example, counting paths made of steps $(1,0)$ and $(0,1)$ under the line $y=\frac12 x$, is equivalent to counting paths in the upper half plane made of steps $(1,1)$ and $(1,-2)$, as shown in Fig. \ref{12path}. Paths of this form are called \emph{excursions} in \cite{banderier200237}. Moreover, counting of these paths is related to the counting of all paths starting at the origin and ending on the $y=0$ line, made of elementary steps $(1,r)$ and $(1,-s)$, and unconstrained (i.e. not constrained to the upper half of the lattice), as shown in Fig.~\ref{12bridge}. Such general paths are called \emph{bridges}, and we denote their generating function by $y_B(x)$. It can be shown that generating functions of excursions and bridges are related by \cite{banderier200237}\footnote{It is also common in literature to take $y(x)=\sum_{k=0}^{\infty} \sum_{\pi\in  \, k\textrm{-paths}} x^{(r+s)k}$ as a generating function of lattice paths, and in such a way the powers of $x$ measure the number of steps (i.e the length) of  a $k$-path. This just reduces to the rescaling of parameter $x$ and consequently an extra factor of $r+s$ in (\ref{bridges}) compared to the formulas in e.g. \cite{banderier200237}. We note that the analogues of the formula (\ref{bridges}) hold for more general paths and bridges, as explained in \cite{banderier200237}.}
\be
y_B(x) = 1 + (r+s)x\frac{d}{dx} \log y_P(x)  = 1 + (r+s)x\frac{y'_P(x)}{y_P(x)}.   \label{bridges}
\ee

The first explicit expression for the generating function $y_P(x)$ in (\ref{Paths-x}) was obtained by Bizley in~\cite{Bizley_paths} and is in fact equivalent to (\ref{bridges}). He proved that
\begin{equation}
  y_P(x) = \exp\Big(\sum_{n=1}^{\infty} g_n x^n \Big),   \label{y-Bizley}
\end{equation}
with the coefficients depending on the slope of the line $y=\frac{r}{s}x$ and expressed through the binomial
\begin{equation}\label{gn}
  g_n = \frac{1}{(r+s)n}\binom{(r+s)n}{r n},
\end{equation}
which is clearly symmetric under the exchange of $r$ and $s$. For example, the number of paths reaching the first point of coordinates $(r,s)$ is
\begin{equation}
  g_1 = \frac{1}{r+s}\binom{r+s}{r}.
\end{equation}
For example, for $r=s=1$ we obtain a classical formula for the numbers of lattice paths under the line $y=x$, which are given by Catalan numbers $C_n=\frac{1}{n+1}\binom{2n}{n}$; i.e. $\sum_{n=0}^\infty C_n x^n=\exp{\big( \sum_{n=1}^{\infty} \binom{2n}{n}\frac{x^n}{2n} \big)}$. For $r=2$ and $s=3$ we find numbers of lattice paths under the line of the slope 2/3
\be
y_P(x) = 1+2 x+23 x^2+377 x^3+7229 x^4+151491 x^5+3361598 x^6 +\ldots    \label{paths-23}
\ee

Note that no straightforward generalization of the Bizley formula is known for the $q$-deformed generating function. However, by invoking the knots-quivers correspondence, in (\ref{qPaths-quiver}) we propose such a $q$-dependent generalization in terms of the quiver data. 

In~\cite{Duchon} Duchon proposed an aproach to the problem of counting paths based on constructing a (noncomuttative) grammar, such that each lattice path corresponds to exactly one word in this grammar. From this perspective the counting of lattice paths is equivalent to counting words. In fact Duchon's approach does not yield directly the generating function $y_P(x)$, but instead it gives an algebraic equation it satisfies
\begin{equation} 
  A(x, y_{P}) = 0. \label{A-poly-paths}
\end{equation}
For example, for the line $y=\frac{2}{3}x$ such an algebraic equation takes form $A(x,y_P) = 1 - y_P + x y_P^5 (2 - y_P + y_P^2) + x^2 y_P^{10}$. This equation can be solved, yielding coefficients in (\ref{Paths-x}). First few of them are listed explicitly in (\ref{paths-23}) and in general, the number of such paths of length $n$ (and for $r=2$, $s=3$) can be written as $c_n(1) = \sum_i \frac{1}{5n + i + 1} \binom{5n+1}{n-i} \binom{5n +2i}{i}$. We present Duchon's formalism in section \ref{ssec-Apoly} in more detail, and then employ it to prove the equivalence of generating functions of knots and lattice paths.

We note that there exist yet another formula for the number of lattice paths $c_n(1)$ for arbitrary $r$ and $s$. In~\cite{banderier200237}, see also~\cite{sato1989}, a unified approach to the lattice paths counting problem was proposed, and the following result was found
\be
 c_{n}(1) = \sum_{\nu_1 + \cdots+ \nu_r = r n} \frac{1}{1+\nu_1 e}\binom{(1+\nu_1 e)/r}{\nu_1} \cdots \frac{1}{1 +\nu_c e } \binom{(1+\nu_c e)/r}{\nu_c}\omega^{\sum_{j=1}^r (j-1)\nu_j}, 
\ee
where  $e= r+s$ and $\omega$ is any $r$-th primitive root of unity. This approach applies more generally, e.g. to paths terminating at a certain height in the upper half plane in fig.~\ref{12path}. 

Finally, let us also mention a relationship between the numbers $g_n$ from (\ref{gn}) and torus knots, which is different from the relationship that we will pursue in this paper. It was shown in \cite{gorsky} that the dimension of the bottom row of the uncolored HOMFLY-PT homology of the $(r,s)$ torus knot 
equals $\frac{1}{r+s}\binom{r+s}{r}$. In particular, the dimension of the bottom row of the uncolored HOMFLY-PT homology of the $(n,n+1)$ torus knot 
equals the $n$-th Catalan number.


\section{Donaldson-Thomas invariants of a symmetric quiver}    \label{sec-DT}

In this section we present the first important result of this paper, namely explicit formulae for invariants of an arbitrary quiver. Such formulas are important in their own right, and to our knowledge they have not been known before. In the rest of the paper we will relate these expressions to knot invariants on one hand, and lattice paths counting on the other hand.

\subsection{Explicit formulae for the classical generating series}\label{sec31}

We provide now general expressions for coefficients $b_{l_1,\ldots,l_m}$ of the classical limit of the generating series (\ref{definv})
\begin{equation}
y(x_1,\ldots,x_m)=\lim_{q\to 1} \frac{P_C(q^2 x_1,\ldots,q^2 x_m)}{P_C(x_1,\ldots,x_m)}=\sum_{l_1,\ldots,l_m} b_{l_1,\ldots,l_m} x_1^{l_1}\ldots x_m^{l_m},   \label{definv-bis}
\end{equation}
where $P_C(x_1,\ldots,x_m)$ is the motivic generating series introduced in (\ref{P-C})
\begin{equation*}
P_C(x_1,\ldots,x_m)=\sum_{d_1,\ldots,d_m} \frac{(-q)^{\sum_{i,j=1}^m C_{i,j}d_i d_j}}{(q^2;q^2)_{d_1}\cdots(q^2;q^2)_{d_m}} x_1^{d_1}\cdots x_m^{d_m},   \label{P-C2}
\end{equation*}
determined by a matrix $C$ of a symmetric quiver with $m$ vertices, i.e. $C$ is an arbitrary symmetric $m\times m$ matrix whose entries are non-negative integers $C_{i,j}$.  

\begin{definition}
Let $k\in\{1,\ldots,m\}$. For a set\footnote{Just to emphasize that here we really mean (an unordered) set, e.g. sets of two pairs $\{(1,2),(1,3)\}$ and $\{(1,3),(1,2)\}$ are considered the same.} of $k$ pairs $(i_u,j_u)$, $u=1,\ldots,k$, where $1\le i_u,j_u\le m$, we say that it is admissible, if it satisfies the following two conditions:\\
(1)\quad  there are no two equal among $j_1,\ldots,j_k$\\
(2)\quad  there is no cycle of any length: for any $l$, $1\le l \le k$, there is no subset of $l$ pairs $(i_{u_{\ell}},j_{u_{\ell}})$, $\ell=1,\ldots,l$, such that $j_{u_{\ell}}=i_{u_{\ell+1}}$, $\ell=1,\ldots,l-1$, and $j_{u_{l}}=i_{u_1}$.
\end{definition}

\begin{proposition}\label{glavna}
Coefficients $b_{l_1,\ldots,l_m}$ in (\ref{definv-bis}) take form
\begin{equation}
b_{l_1,\ldots,l_m} = (-1)^{\sum_{i=1}^m (C_{i,i}+1) l_i} A(l_1,\ldots,l_m)
\prod_{j=1}^m\frac{1}{1+\sum_{i=1}^m{C_{i,j}l_i}} \left({1+\sum_{i=1}^m{C_{i,j}l_i}} \atop {l_j}  \right)    \label{b1-glavna}
\end{equation}
where
\begin{equation}\label{forA1}
A(l_1,\ldots,l_m)= 1+\sum_{k=1}^{m-1}\sum_{admissible\, \Sigma_k}\prod_{(i_u,j_u)\in\Sigma_k} C_{i_u,j_u} l_{i_u}.   
\end{equation}
Here, in the second sum, we are summing over all admissible subsets of length $k$ -- one such subset we denote $\Sigma_k$. Note that $A(l_1,\ldots,l_m)$ is a polynomial in variables $l_i$, of degree $m-1$, whose coefficients depend only on the off-diagonal entries of $C$.
\end{proposition}

This proposition can be proven by induction, generalizing the results that we found for the matrix $C$ of size $m=2$ or $m=3$. To this end it is useful to take advantage of an alternative definition of $A(l_1,\ldots,l_m)$, which is not as explicit as (\ref{forA1}), but rather involves an induction on $m$. Namely, for a given matrix $C=[C_{i,j}]_{i,j=1}^m$, we shall define a certain polynomial in $m$ variables, $P_m(C)(x_1,\ldots,x_m)$, whose coefficients are sums and multiples of the entries of $C$, with specific properties. In order to state those properties, we define first an action of the permutation group $S_m$ on $m\times m$ matrices, and on the polynomials of the form $p(C)(x_1,\ldots,x_m)$, whose coefficients are functions of the entries of $C$. For a permutation $\sigma\in S_m$ we define its action on $m \times m$ matrices as follows:
\begin{equation}
\left[\sigma \circ C\right]_{i,j} := C_{\sigma_i,\sigma_j}, \quad\quad i,j=1,\ldots,m,
\end{equation} 
and on polynomials $p(C)(x_1,\ldots,x_m)$ by
\begin{equation}
\sigma \circ p(C)(x_1,\ldots,x_m):=p(\sigma\circ C)(x_{\sigma_1},\ldots,x_{\sigma_m}).
\end{equation} 
The first property that we require on $P_m$'s is that they are invariant under the action of $S_m$
\begin{equation}
(\textrm{A1})\quad\quad  \sigma  \circ P_m(C)(x_1,\ldots,x_m)=   P_m(C)(x_1,\ldots,x_m),\quad \forall\sigma \in S_m,
\end{equation}
and the second property is an inductive one
\begin{equation}
(\textrm{A2})\quad\quad  P_m(C)(x_1,\ldots,x_{m-1},0)=   P_{m-1}(C')(x_1,\ldots,x_{m-1}) \cdot\Big(1+\sum\limits_{i=1}^{m-1} C_{i,m} x_i \Big),
\end{equation}
where $C'$ denotes the submatrix of $C$ formed by its first $m-1$ rows and columns. These two properties, together with the initial condition
\begin{equation}
(\textrm{A0})\quad\quad  P_1(C)(x)=1,
\end{equation}
uniquely determine $P_m(C)(x_1,\ldots,x_m)$. Then, for a given $m\times m$ matrix $C$, the alternative description of $A(l_1,\ldots,l_m)$ from (\ref{forA1}) simply reads
\begin{equation}
A(l_1,\ldots,l_m)=P_m(C)(l_1,\ldots,l_m).
\end{equation}

\subsubsection*{Examples for small $m$}

It is useful to present explicit expressions for $b_{l_1,\ldots,l_m}$ for several small values of $m$. First, for $m=1$, we consider a quiver that consists of a single vertex and $f\in \Z_{\ge 0}$ loops, whose structure is encoded in the matrix
\begin{equation}\label{1vert}
C = \left[ f \right].
\end{equation}
In this case the coefficients ${b}_{i}(1)$ in (\ref{b1-glavna}) are given by 
\begin{equation} 
{b}_{i} = \frac{(-1)^{(f+1) i}}{f i +1}\left({f  i+1 \atop i}\right).     \label{1vertcoeffb}
\end{equation}

Now, consider a quiver with $m=2$ vertices, determined by an arbitrary $2\times 2$ symmetric matrix
\begin{equation}\label{2vert}
C=\left[\begin{array}{cc}
\alpha & \beta \\
\beta & \gamma
\end{array}
\right],
\end{equation}
where $\alpha$, $\beta$ and $\gamma$ are arbitrary non-negative integers. In this case coefficients ${b}_{i,j}(1)$ in (\ref{b1-glavna}) are given by 
\begin{equation}
{b}_{i,j} = (-1)^{(\alpha+1) i + (\gamma+1) j}\frac{\beta i+\beta j+1}{(\alpha i+\beta j+1)(\beta i+\gamma j+1)}\left({\alpha i+\beta j+1 \atop i}\right)\left({\beta i+\gamma j+1 \atop j}\right).   \label{C2x2-b-ij}
\end{equation}
Furthermore, consider a quiver with $m=3$ vertices, determined by an arbitrary $3\times 3$ symmetric matrix
\begin{equation}\label{3vert}
C=\left[\begin{array}{ccc}
\alpha & \beta &\delta \\
\beta & \gamma&\epsilon\\
\delta&\epsilon&\phi
\end{array}
\right],
\end{equation}
where $\alpha$, $\beta$, $\gamma$, $\delta,$ $\epsilon$, and $\phi$ are arbitrary non-negative integers. In this case coefficients ${b}_{i,j,k}(1)$ in (\ref{b1-glavna}) are given by 
\begin{align}
  {b}_{i,j,k} =&  (-1)^{(\alpha+1) i + (\gamma+1) j + (\phi+1) k} A_{i,j,k}\left({\alpha i+\beta j+\delta k+1 \atop i}\right)\left({\beta i+\gamma j+ \epsilon k+1 \atop j}\right)\nonumber \\
  &\times \left({\delta i+\epsilon j+ \phi k+1 \atop k}\right),
\end{align}
where 
\be
\begin{split}
A_{i,j,k}=&\frac{1}{(\alpha i+\beta j+\delta k+1)(\beta i+\gamma j+\epsilon k+1)(\delta i+\epsilon j+\phi k +1)}\times \\
&\times \big(1+(\beta +\delta) i+(\beta+\epsilon) j+(\delta+\epsilon)k+ \beta\delta i^2+
\beta\epsilon j^2+\delta\epsilon k^2+\\
&\qquad +\beta(\delta+\epsilon)ij+ \delta(\beta+\epsilon)ik+\epsilon(\beta+\delta)jk \big).
\end{split}
\ee


\subsection{Explicit formulae for Donaldson-Thomas invariants}

We now determine explicitly numerical Donaldson-Thomas invariants $\Omega_{d_1, \dots, d_m}$ of an arbitrary symmetric quiver. Recall that they are defined by the factorization in (\ref{definv-class})
\begin{equation}
\begin{split}
y(x_1,\ldots,x_m)&=\lim_{q\to 1}\frac{P_C(q^2 x_1,\ldots,q^2 x_m)}{P_C(x_1,\ldots,x_m)}=\sum_{l_1,\ldots,l_m} b_{l_1,\ldots,l_m} x_1^{l_1}\ldots x_m^{l_m} = \\
&= \prod_{(d_1,\ldots,d_m)\neq 0} \big(1 -  x_1^{d_1}\cdots x_m^{d_m}  \big)^{\Omega_{d_1, \dots, d_m}}.   \label{defy}
\end{split}
\end{equation}
Donaldson-Thomas invariants $\Omega_{d_1, \dots, d_m}$ can be easily extracted from the logarithmic derivative of the function $y(x_1,\ldots,x_m)$. Therefore the crucial task is to determine the logarithm of $y(x_1,\ldots,x_m)$. We find that it is given by an expression closely related to $b_{l_1,\ldots,l_m}$ in (\ref{b1-glavna}).

\begin{proposition}\label{glavnalog}
The logarithm of $y(x_1,\ldots,x_m)$ in (\ref{defy}) takes form
\begin{align}\label{forlog}
\log y (x_1,\ldots,x_m)= & \sum_{{\scriptsize{\begin{array}{c}l_1,\ldots,l_m\ge 0\\
                                             l_1+\cdots+l_m>0\end{array}}}}  (-1)^{\sum_{i=1}^m (C_{i,i}+1) l_i} A_{\max}(l_1,\ldots,l_m) \times \nonumber \\
&\times \prod_{j=1}^m\frac{1}{\sum_{i=1}^m{C_{i,j}l_i}} \left({\sum_{i=1}^m{C_{i,j}l_i}} \atop {l_j}  \right)   x_1^{l_1}\cdots x_m^{l_m},
\end{align}
where
\begin{equation}\label{forAmax1}
A_{\max}(l_1,\ldots,l_m)= \sum_{admissible\, \Sigma_{m-1}}\prod_{(i_u,j_u)\in\Sigma_{m-1}} C_{i_u,j_u} l_{i_u}.  
\end{equation}
Here, in (\ref{forAmax1}), we sum over all admissible subsets $\Sigma_{m-1}$ of length $m-1$, which is in fact the maximal possible length of an admissible set. In other words, the factor $A_{\max}(l_1,\ldots,l_m)$ is the homogeneous part, of the top-degree, of the polynomial $A(l_1,\ldots,l_m)$ in (\ref{forA1}).
\end{proposition}

Again, we have an alternative, inductive definition for $A_{\max}$. As in section \ref{sec31}, for an $m\times m$ matrix $C$ we define polynomials $P_m^{\max}(C)(x_1,\ldots,x_m)$ in $m$ variables with coefficients being sums and products of $C_{i,j}$'s. The action of the permutation $\sigma\in S_m$ on matrices $C$ and polynomials $P_m^{\max}$ is defined in the same way as in section \ref{sec31}. Then we require:
\begin{eqnarray}
&(\textrm{A0'})&\quad\quad P_1^{\max}(x)=1,\\
&(\textrm{A1'})&\quad\quad \sigma \circ P_m^{\max}(C)(x_1,\ldots,x_m)=P_m^{\max}(C)(x_1,\ldots,x_m), \quad \forall \sigma\in S_m,\\
&(\textrm{A2'})&\quad\quad P_m^{\max}(C)(x_1,\ldots,x_{m-1},0)=  P_{m-1}^{\max}(C')(x_1,\ldots,x_{m-1}) \sum\limits_{i=1}^{m-1} C_{i,m} x_i,
\end{eqnarray}  
where $C'$ is obtained from $C$ by erasing its last row and column. These three axioms uniquely determine $P_m^{\max}(C)(x_1,\ldots,x_m)$. Then we have an alternative description for $A_{\max}(l_1,\ldots,l_m)$
\begin{equation}\label{forAmax2}
A_{\max}(l_1,\ldots,l_m)=P_m^{\max}(C)(l_1,\ldots,l_m).
\end{equation}
The above proposition can be proven by induction on $m$. 

\subsubsection*{Examples for small $m$}

It is again instructive to present explicitly examples for some values of $m$. Consider first the simplest case of $m=1$, i.e. a quiver with a single vertex and $f$ loops, so that
$$
C=\left[ f \right].
$$
Then the formula (\ref{forlog}) reduces to
\begin{equation}\label{for1}
(\log y)(x)=\sum_{n > 0}  \frac{(-1)^{(f+1)n}}{f n} \left(f n \atop n  \right) x^n.
\end{equation}

For $m=2$, in the case of quivers with two vertices defined by $2\times 2$ symmetric matrices
\begin{equation}\label{2vert2}
C=\left[\begin{array}{cc}
\alpha & \beta \\
\beta & \gamma
\end{array}
\right],
\end{equation}
where $\alpha$, $\beta$ and $\gamma$ are arbitrary non-negative integers, the formula (\ref{forlog}) becomes
\begin{equation}
\log y(x_1,x_2)=\sum_{{\scriptsize{\begin{array}{c}i,j\ge 0\\
i+j>0\end{array}}}} 
(-1)^{(\alpha+1)i + (\gamma+1)j}\frac{\beta i+\beta j}{(\alpha i+\beta j)(\beta i+\gamma j)}\left({\alpha i+\beta j \atop i}\right)\left({\beta i+\gamma j \atop j}\right) x_1^i x_2^j.    \label{log-y-x1x2}
\end{equation}
It follows that Donaldson-Thomas invariants take form
\begin{align}
  \Omega_{r,s}=&\; -\frac{1}{(r+s)}\sum_{d|\gcd(r,s)} (-1)^{(\alpha+1)r/d + (\gamma+1)s/d} \mu (d) 
                 \frac{(r/d+s/d)(\beta r/d+\beta s/d)}{(\alpha r/d+\beta s/d)(\beta r/d +\gamma s/d)}  \times \nonumber \\
               &\times\left({\alpha r/d+\beta s/d \atop r/d}\right)\left({\beta r/d+\gamma s/d \atop s/d}\right) = \nonumber \\
  =&\;  -\frac{\beta (r+s)}{(\alpha r+\beta s)(\beta r+\gamma s)}\sum_{d|\gcd(r,s)} (-1)^{(\alpha+1)r/d + (\gamma+1)s/d}
     \mu(d)  \times \nonumber \\
               &\times\left({\alpha r/d+\beta s/d \atop r/d}\right) \left({\beta r/d+\gamma s/d \atop s/d}\right), 
\end{align}
for all $(r,s)\in\N^2\setminus\{(0,0)\}$. Integrality of these invariants implies that $(\alpha r+\beta s)(\beta r+\gamma s)$ in the denominator above divides the rest of the expression, which is a nontrivial number theoretic prediction. Furthermore, specializing to diagonal invariants (\ref{nr-DT}), we find that 
\begin{equation}
n_r = \frac{1}{r^2}\sum_{d|r} \mu (\frac{r}{d}) \sum_{i+j=d} (-1)^{(\alpha+1)i + (\gamma+1)j}
\frac{(i+j)(\beta i+\beta j)}{(\alpha i+\beta j)(\beta i +\gamma j)}\left({\alpha i+\beta j \atop i}\right)\left({\beta i+\gamma j \atop j}\right) \in\N,  \label{n-r-2x2}
\end{equation}
for all $r\in\N$, i.e. $r^2$ divides the rest of the above expression, which is also an interesting divisibility property.

For $m=3$ and an arbitrary $3\times 3$ symmetric matrix
\begin{equation}\label{3vert-2}
C=\left[\begin{array}{ccc}
\alpha & \beta &\delta \\
\beta & \gamma&\epsilon\\
\delta&\epsilon&\phi
\end{array}
\right],
\end{equation}
where $\alpha$, $\beta$, $\gamma$, $\delta,$ $\epsilon,$ and $\phi$ are arbitrary non-negative integers, the formula (\ref{forlog}) reduces to
\begin{align}
\log y (x_1,x_2,x_3)=&\;\sum_{{\scriptsize{\begin{array}{c}i,j,k\ge 0\\
                                          i+j+k\!>\!0\end{array}}}} (-1)^{(\alpha+1)i + (\gamma+1)j + (\phi+1)k} A^{\max}_{i,j,k} \times \nonumber \\
  &\times\left({\alpha i+\beta j+\delta k \atop i}\right)\left({\beta i+\gamma j+ \epsilon k\atop j}\right)\left({\delta i+\epsilon j+ \phi k \atop k}\right) x_1^i x_2^j x_3^k,
\end{align}
 where 
\be
A^{\max}_{i,j,k}=\frac{ \beta\delta i^2+\beta(\delta+\epsilon)ij+
\beta\epsilon j^2+ \delta(\beta+\epsilon)ik+\epsilon(\beta+\delta)jk
+\delta\epsilon k^2}{(\alpha i+\beta j+\delta k)(\beta i+\gamma j+\epsilon k)(\delta i+\epsilon j+\phi k )}.
\ee
In this case, for diagonal invariants (\ref{nr-DT}) we find
\begin{align}
n_r =& \frac{1}{r^2} \sum_{d|r}  \mu (\frac{r}{d})
  \sum_{i+j+k=d} (-1)^{(\alpha+1)i + (\gamma+1)j + (\phi+1)k} (i+j+k)A^{\max}_{i,j,k} \times \nonumber \\
  &\times \left({\alpha i+\beta j+\delta k \atop i}\right)\left({\beta i+\gamma j+ \epsilon k\atop j}\right)\left({\delta i+\epsilon j+ \phi k \atop k}\right).  \label{n-r-3x3} 
\end{align}
Because $n_r \in \mathbb{N}$, it is also an interesting property of divisibility by $r^2$.

Note that BPS numbers $n_r$ for knots, i.e. diagonal invariants (such as (\ref{n-r-2x2}) and (\ref{n-r-3x3})) for quivers that are associated to knots, are divisible by an additional factor of $r$, i.e. $\frac{n_r}{r}\in\N$, as found in \cite{Garoufalidis:2015ewa}. On the other hand, by considering many examples of quivers associated to random (symmetric) matrices $C$ we realized that such an extended divisibility does not hold in general. This confirms that invariants associated to knots are in some way special; it is desirable to understand precise origin of these special properties.


\section{Torus knots and counting paths}    \label{sec-knots-paths}

Having introduced all necessary ingredients, in the following proposition we state the second main result of this work. In this proposition by extremal invariants we mean maximal (top row) invariants of left-handed torus knots, as explained at the end of section \ref{ssec-knots-quivers}.

\begin{proposition}  \label{prop2}
The generating function $y_P(x)$ in (\ref{Paths-x}) of lattice paths under the line of the slope $r/s$ is equal to the classical generating function (\ref{yxa-extremal}) of maximal (top row) HOMFLY-PT invariants of the left-handed $(r,s)$ torus knot in framing $rs$. That is
\be
y_P(x) = \sum_{k=0}^{\infty} \sum_{\pi\in  \, k\textrm{-paths}} x^k = \lim_{q\to 1} \frac{P^{+}(q^2\bar{x})}{P^{+}(\bar{x})},   \label{yPx-prop}
\ee
where $P^{+}(\bar{x})$ is the generating series of maximal HOMFLY-PT polynomials defined in (\ref{Pr-LMOV-minmax}), which can be also expressed in terms of the corresponding quiver (\ref{PxC-ext}). The variables $x$ and $\bar{x}$ are related through
\be
 x = (-1)^{\sum_i (t_i+1)} \bar{x},
 \ee
 where $t_i$'s are homological degrees of the torus knot $(r,s)$ in framing $rs$. 
\end{proposition}

This statement has further consequences. First, it follows that algebraic equations satisfied by these generating functions -- i.e. (extremal) A-polynomials and equations determined by the Duchon grammar -- are the same. Second, via the knots-quivers correspondence, the generating function of lattice paths (\ref{Paths-x}) can be expressed in terms of diagonal quiver invariants (\ref{yx-Bn}), which are combinations of classical quiver invariants $b_{l_1,\ldots,l_m}\equiv b_{l_1,\ldots,l_m}(1)$ defined in (\ref{definv-class}), which we determined explicitly in (\ref{b1-glavna}). The invariants $b_{l_1,\ldots,l_m}$ are interesting in themselves and provide a refinement of numbers of lattice paths; they should also have a natural combinatorial interpretation as counting some particular paths. 

Proposition \ref{prop2} also implies an interesting relation of classical LMOV invariants (\ref{n-r}), or diagonal DT invariants (\ref{nr-DT}), to the counting functions of bridges (\ref{bridges}). Indeed, note that classical LMOV and diagonal DT invariants are encoded in the logarithmic derivative (\ref{log-y}). Similarly, the generating function of bridges is also given by (1 plus) the logarithmic derivative (\ref{bridges}). It follows that invariants $n_r$ in (\ref{n-r}) or (\ref{nr-DT}) are expressed as combinations of the numbers of bridges, divided by $r^2$. This also means that these combinations of the numbers of bridges are divisible by $r^2$, which is quite a nontrivial statement; it would be interesting to find its combinatorial interpretation. 


Furthermore, it is natural to expect that there exists a quantum deformation of Proposition \ref{prop2}. The parameter $q$ that computes the area under lattice paths, as well as the parameter $q$ of the HOMFLY-PT polynomial, are two parameters that provide natural deformations of the generating functions that we consider. However, it turns out that quantum deformations associated to these two parameters are dif\mbox{}ferent, and in order to find agreement of $q$-deformed generating functions of paths and knot polynomials, some adjustment is necessary. Amusingly, we find that $q$-weighted paths are encoded in the quiver generating function, with appropriate identification of each $x_i$ with $x$ (which is different than such an identification for knots). In the following proposition we present explicit formulae for such $q$-deformed path counting (which is also closely related to extremal invariants of torus knots, the only difference being a different identification of parameters $x_i$).

\begin{proposition} \label{prop2-q}
The generating function $y_{qP}(x)$ in (\ref{yqPx})  of lattice paths under the line of the slope $r/s$, weighted by the area between this line and a given path, is equal to the following ratio of quiver motivic generating functions $P_C(x_1,\ldots,x_m)$ introduced in (\ref{P-C}), with identification of parameters $x_i =  (-1)^{C_{i,i}+1} q^{-1} x$
\be
y_{qP}(x) = \sum_{k=0}^{\infty} \sum_{\pi\in\,  k\textrm{-paths}} q^{area(\pi)}  x^k = \frac{P_C(q^2x_1,\ldots,q^2 x_m)}{P_C(x_1,\ldots,x_m)}\Big|_{x_i = x q^{-1}}  . \label{qPaths-quiver}
\end{equation}
For the line of the slope $r/s$, the quiver in question is defined by the matrix $C$ that encodes maximal invariants of left-handed $(r,s)$ torus knot in framing $rs$. The coefficients $b_{l_1, \dots, l_m}^{(r,s)}$ appearing in the expansion of the classical generating series $y_P(x)$ are related to the corresponding coefficients $b_{l_1, \dots, l_m}$ of the expansion of quiver motivic function through
\begin{equation}
  b_{l_1, \dots, l_m}^{(r,s)} = (-1)^{\sum_{i=1}^m (C_{i,i} + 1)l_i} b_{l_1, \dots, l_m}.
\end{equation}
\end{proposition}

In the rest of this section we illustrate the relation between the lattice paths, invariants of torus knots and quivers in various examples. 


\subsection{Unknot and Fuss-Catalan numbers}   \label{ssec-unknot}

As a warm up, let us consider a framed unknot. For framing $f$, it can also be thought of as $(f,1)$ torus knot. The minimal colored HOMFLY-PT polynomial of the unknot, in the trivial ($f=0$) framing, reads
\be
P_r^-(q) = \frac{q^r}{(q^2;q^2)_r},
\ee
so that the generating function of minimal invariants for the framed unknot takes form
\be
P^-(x) = \sum_{r\geq 0} x^r q^{fr(r-1)} P_r^-(q) = \sum_{r\geq 0} x^r q^{r(1-f)} \frac{q^{fr^2}}{(q^2;q^2)_r}.
\ee
This generating function is simply related to the quiver generating function (\ref{P-C}), for a quiver with one vertex and $f$ loops 
\be
P^-(x) = P_C((-1)^f q^{1-f} x ), \qquad C = \big[ f \big].
\ee
The coefficients of the classical generating series $y_P(x)$ in (\ref{yPx-prop}) in this case are related to the coefficients $b_i$ in (\ref{1vertcoeffb}) through  
\be
b_i^{(f,1)} = (-1)^{(f+1)i} b_i = \frac{1}{f i + 1}\binom{fi + 1}{i},
\ee
and are specializations of Fuss-Catalan numbers, which are indeed known to count lattice paths under the line $y = f x$. Therefore the $f$-framed unknot is related to the lattice paths under the $y= f x$ line. The corresponding BPS number are given by the formula~\eqref{for1}. Similarly, the $q$-weighted paths are given by $q$-deformed Fuss-Catalan numbers, and their generating function is given by (\ref{qPaths-quiver}). The relation between the unknot invariants, Fuss-Catalan numbers, Donaldson-Thomas invariants for $f$-loop quiver, and LMOV invariants have been also considered also in \cite{Rei12,Kucharski:2016rlb}. We generalized results briefly summarized here to the full $a$-dependent unknot invariants -- which turn out to correspond to so-called Schr{\"o}der paths -- in section \ref{sec-schroeder}.


\subsection{BPS numbers from Bizley formula}

As another simple application of the paths/torus knots correspondence we consider the extremal invariants $n_p$ defined in~\eqref{n-r} as
\be
n_p = \frac{1}{p^2} \sum_{d|p} \mu(d) a_{\frac{p}{d}},
\ee
with the M\"obius function $\mu(d)$. The coefficients $a_k$ are related to the logarithmic derivative of $y(x)$
\be
 x \frac{d}{dx}\log y(x) = \sum_{k=0}^{\infty} a_k x^k.
\ee
Under the paths/torus knots correspondence the function $y(x)$ of the $(r,s)$ torus knot in framing $rs$ is related to the counting function of lattice paths under the line of slope $r/s$. The only subtle point is a proper change of variables, which takes form
\be
y_P(x) = y((-1)^{t_i+1} x),
\ee
where $t_i$'s are homological degrees.  
The counting function $y_P(x)$ is given by the Bizley formula~\eqref{y-Bizley}. Straightforward computation gives
\be
a_k = \frac{(-1)^{p\sum_i (t_{i} + 1)}}{r+s}\binom{(r+s)k}{rk}.
\ee
The homological degrees $t_i$ are equal to the diagonal entries of the quiver matrix $C$.
In the Table~\ref{tab-BPS} we give the BPS numbers $n_p$ for several torus knots.
\begin{table}[h]
\begin{small}
\be
\begin{array}{|c|c|}
\hline 
\textrm{\bf Torus knot} & n_p   \nonumber \\
\hline 
\hline
(1,1) & -1,\, 1,\, -1,\, 2,\, -5,\, 13,\, -35,\, 100,\, -300, \dots \\
\hline
(1,2) & 1,\, 1,\, 3,\, 10,\, 40,\, 171,\, 791,\, 3828,\, 19287, \dots \\
\hline
(2,3) &  2,\, 10,\, 111,\, 1572,\, 26150,\, 480489,\, 9469222, \dots \\
\hline
(2,5) &  3,\, 35,\, 861,\, 27742,\, 1049025,\, 43881197, \dots \\
\hline
(2,7) &  4,\, 84,\, 3654,\, 210120,\, 14178610,\, 1058662314, \dots \\
\hline
(3,4) &  5,\, 106,\, 4665,\, 271596,\, 18559675,\, 1403558826, \dots\\
\hline
(3,5) &\  -7,\, 252,\, -18159,\, 1763944,\, -201126725,\, 25381382988, \dots\ \\
\hline
\end{array}
\ee  
\end{small}
\caption{BPS numbers $n_p$ for torus knots obtained from the paths/torus knots correspondence and the Bizley formula.} \label{tab-BPS}
\end{table}


\subsection{Path counting and BPS numbers from quivers}   \label{ssec-paths-quivers}

We illustrate now in more involved examples how explicit expressions for quiver generating functions and corresponding BPS invariants provide new (or reproduce known) expressions for counting paths. Recall first that quivers for maximal invariants for family of $(2,2p+1)$ left-handed torus knots, in framing $2(2p+1)$ (recall conventions presented at the end of section \ref{ssec-knots-quivers}), take form \cite{Kucharski:2017ogk} 
\begin{equation}
C=\left[
\begin{array}{ccccc}
2(2p+1)+1 & 2(2p+1)-1 & 2(2p+1)-3 & \cdots & 2(2p+1)+1-2p\\
2(2p+1)-1 & 2(2p+1)-1 & 2(2p+1)-3 & \cdots & 2(2p+1)+1-2p\\
2(2p+1)-3 & 2(2p+1)-3 & 2(2p+1)-3 & \cdots & 2(2p+1)+1-2p\\
\vdots  & \vdots & \vdots & \ddots & \vdots\\
2(2p+1)+1-2p & 2(2p+1)+1-2p & 2(2p+1)+1-2p & \cdots & 2(2p+1)+1-2p
\end{array}    \label{C-22p1}
\right]
\end{equation}
For example, for trefoil, $(2,5)$ and $(2,7)$ torus knots, corresponding respectively to $p=1,2,3$, we get
\arraycolsep 3pt
\begin{equation}\label{CT23}
C^{(2,3)}=\left[
\begin{array}{cc}
7 & 5\\
5 & 5
\end{array}
\right], 
\qquad C^{(2,5)}=\left[
\begin{array}{ccc}
11 & 9 & 7\\
9  & 9  & 7\\
7 & 7 & 7
\end{array}
\right], \qquad C^{(2,7)}=\left[
\begin{array}{cccc}
15 & 13 & 11 & 9\\
13  & 13 & 11 & 9\\
11 & 11 & 11 & 9\\
9 & 9 & 9 & 9
\end{array}
\right].
\end{equation}

\begin{figure}[h]
\centering






\includegraphics[scale=0.3]{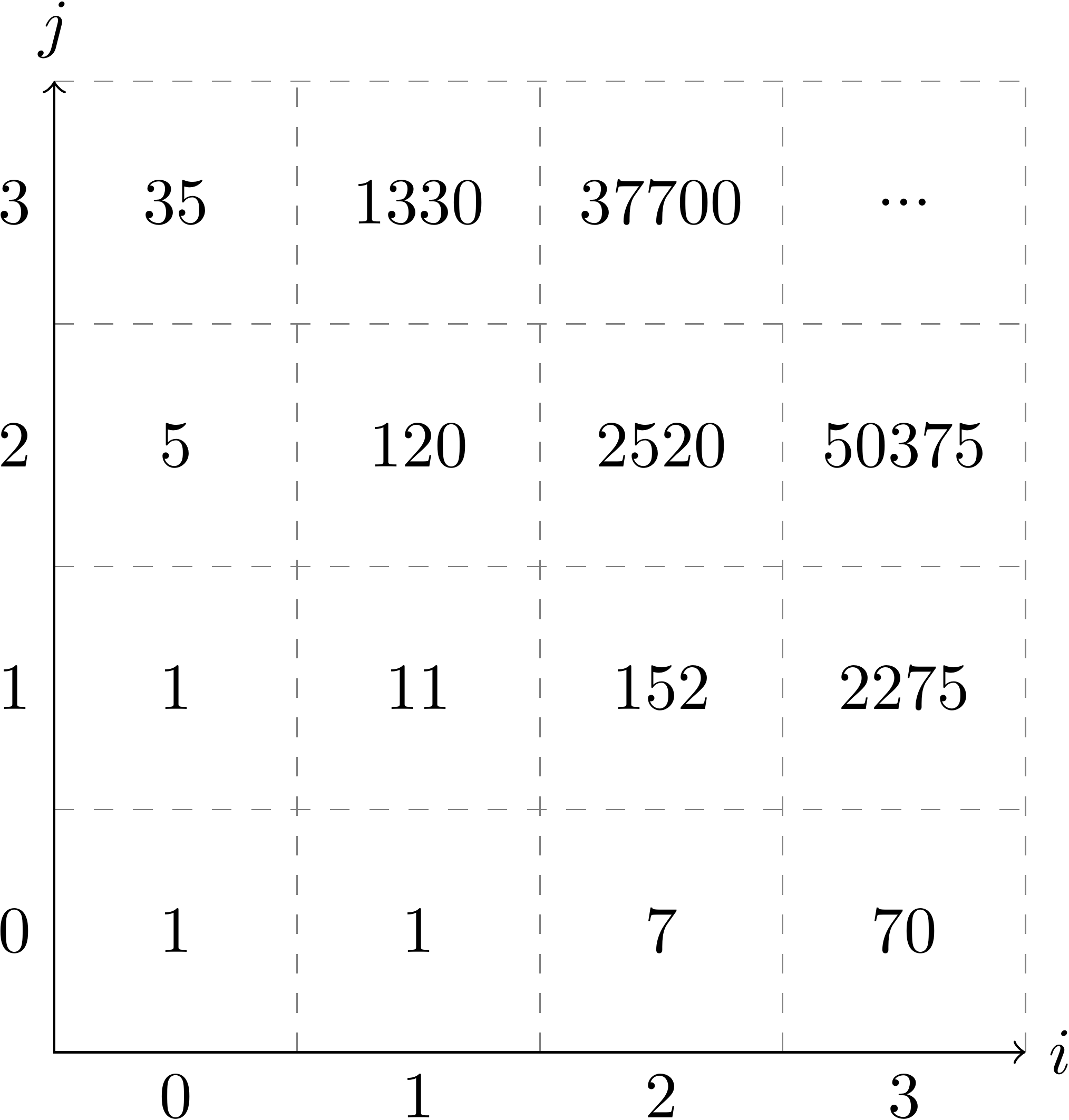}
\caption{Coefficients $b^{(2,3)}_{i,j}$ that refine enumeration of paths under the line of the slope 2/3. Diagonal combinations of these numbers reproduce coefficients in (\protect\ref{paths-23-bis}).}   \label{bij-trefoil}
\end{figure}

Let us consider in detail the trefoil knot, i.e. $(2,3)$ torus knot. The generating function of paths under the line of the slope 2/3 can be computed by the relation to colored extremal knot polynomials via (\ref{yPx-prop}) and it agrees with the outcome of the Bizley formula (\ref{paths-23})
\be
y_P(x) = 1+2 x+23 x^2+377 x^3+7229 x^4+151491 x^5+3361598 x^6 +\ldots    \label{paths-23-bis}
\ee
Furthermore, once the trefoil quiver in (\ref{CT23}) is identified, we can naturally produce the classical quiver generating function (\ref{definv-bis}) that in this case depends on two parameters 
\be
y(x_1,x_2) = \sum_{i,j} b^{(2,3)}_{i,j} x_1^{i} x_2^{j}.
\ee
The coefficients $b^{(2,3)}_{i,j}$, given by (\ref{b1-glavna}), take the explicit form that follows from (\ref{C2x2-b-ij})
\begin{equation}\label{DT1}
b^{(2,3)}_{i,j}=\frac{1}{7i+5j+1}\left({7i+5j+1 \atop i}\right)\left({5i+5j+1 \atop j}\right),
\end{equation}
and we present a few of them in fig. \ref{bij-trefoil}. It is desirable to understand combinatorial interpretation of these numbers. Of course, their diagonal combinations agree with coefficients in (\ref{paths-23-bis}). In fact, these diagonal combinations reproduce the original Duchon formula for the number of lattice paths under the line of the slope 2/3
\begin{eqnarray}\label{duch}
B^{(2,3)}_n&=&\sum_{i+j=n}\frac{1}{7i+5j+1}\left({7i+5j+1 \atop i}\right)\left({5i+5j+1 \atop j}\right)  =  \\
&=& \sum_{i=0}^n \frac{1}{5n+i+1}\left({5n+2i \atop i}\right)\left({5n+1 \atop n-i}\right).
\end{eqnarray}

We can also consider Donaldson-Thomas invariants, which -- as explained above -- are closely related to the number of bridges (\ref{bridges}). For the generating function $y(x) = \sum_n B^{(2,3)}_n x^n$, from (\ref{log-y-x1x2}) we get
\begin{equation}\label{for414}
(\log y)(x)=\sum_{{\scriptsize{\begin{array}{c}i,j\ge 0\\
i+j>0\end{array}}}}
\frac{1}{7 i+5 j}\left({7 i+5 j \atop i}\right)\left({5 i+5 j \atop j}\right) x^{i+j},
\end{equation}
which implies that
\begin{equation}
\frac{x y'}{y}=x (\log y)'=\sum_{n\ge 1}\sum_{i+j=n}
\frac{i+j}{7 i+5 j}\left({7 i+5 j \atop i}\right)\left({5 i+5 j \atop j}\right) x^{n}.
\end{equation}
It then follows from (\ref{n-r}) that extremal, classical BPS numbers for trefoil take form
\begin{equation}
n_r=\frac{1}{r^2}\sum_{d|r} \mu \left(\frac{r}{d}\right) \sum_{i+j=d}
\frac{i+j}{7 i+5 j}\left({7 i+5 j \atop i}\right)\left({5 i+5 j \atop j}\right),
\end{equation}
where $\mu(d)$ is the M\"obius function. One can check that these $n_r$ are indeed integer, as predicted by the LMOV conjecture.

Moreover, in this particular case by using the relationship between the generating function of paths and bridges (\ref{bridges}) for the $(2,-3)$ paths, from (\ref{for414}) we rediscover the identity
\begin{equation}
\left({5n \atop 2n}\right)=\sum_{i=0}^n
\frac{5n}{5n+2i}\left({5n+2i \atop i}\right)\left({5n \atop n-i}\right).     \label{ex-bridge-23}
\end{equation}
However, in general, for a generic quiver, or arbitrarily framed torus (or non-torus) knot, we do not find such simplification.
 
Finally we discuss $q$-weighted path counting. From (\ref{qPaths-quiver}) we find that the $q$-weighted generating function of paths takes form
\begin{align}
y_{qP}(x) =&\,  1 + (q^4 + q^6) x + (q^8+3 q^{10}+4 q^{12}+4 q^{14}+4 q^{16}+3 q^{18}+2 q^{20}+q^{22}+q^{24}) x^2 +  \nonumber \\
&+ (q^{12}+5 q^{14}+12 q^{16}+20 q^{18}+28 q^{20}+34 q^{22}+37 q^{24}+37 q^{26}+36 q^{28} +  \nonumber \\ 
& +33 q^{30}+29 q^{32}+25 q^{34}+21 q^{36}+17 q^{38}+13 q^{40}+10 q^{42}+7 q^{44}+5 q^{46} +  \nonumber \\ 
& +3 q^{48}+2 q^{50}+q^{52}+q^{54}) x^3 + \dots
\end{align}
It is immediate to check that powers of $q$ in this expression indeed compute the area between the line of the slope 2/3 and a given path, analogously as in fig. \ref{14path}. For $q=1$ this expression reduces to (\ref{paths-23-bis}).

The above results generalize to all $p$. For example for $p=2$, i.e. $(2,5)$ torus knot, we find that the number of lattice paths under the line $y=\frac{2}{5}x$, from $(0,0)$ to $(5n,2n)$ is equal
\begin{align}\label{duch25}
\begin{split}
B^{(2,5)}_n&=\sum_{i+j+k=n}\frac{1}{11i+9j+7k+1}\times \\
&\qquad \times \left({11i+9j+7k+1 \atop i}\right)\left({9i+9j+7k+1 \atop j}\right)  \left({7i+7j+7k+1 \atop k}\right).
\end{split}
\end{align}
 
In general, for every $p\ge 1$, we get that the number of directed lattice paths from $(0,0)$ to $(2p+1)n, 2n)$, that stay below the line $y=\frac{2}{2p+1} x$, is given by
\begin{equation}\label{duch2p}
B^{(2,2p+1)}_n=\sum_{i_1+\ldots+i_{p+1}=n}\frac{1}{1+\sum_{j=1}^{p+1} (4p+5-2j) i_j }\,\,\prod_{j=1}^{p+1} \left({1+\sum_{l=1}^{p+1} (4p+5-2\max{(j,l)}) i_l\atop i_j}\right).
\end{equation} 


\subsection{Path counting and BPS numbers for $(3,4)$ torus knot}

Let us consider now another, more involved example of $(3,4)$ torus knot, which corresponds to counting of paths under the line of the slope $\frac{3}{4}$. The (extremal) quiver for $(3,4)$ torus knot was found in \cite{Kucharski:2017ogk}, and for maximal invariants in framing 12, for its left-handed version, it reads
\begin{equation} \label{quiver34}
C^{(3,4)}=\left[\begin{array}{ccccc}
7\, &7&7&7&7\\
7\, &9&8&9&9\\
7\, &8&9&9&10\\
7\, &9&9&11&11\\
7\, &9&10&11&13\end{array}\right]
\end{equation}
From Proposition \ref{glavna} we first determine $b_{l_1,\ldots,l_5}(1)$ as a function of $l_i$. The main part of the expression for ${b_{l_1,\ldots,l_5}(1)}$ is $A(l_1,\ldots,l_5)$ in (\ref{forA1}), which for the above quiver $C^{(3,4)}$ takes form
\begin{eqnarray*}\label{34explicit}
\scriptstyle{A_{(3,4)}(l_1,l_2,l_3,l_4,l_5)} &&\scriptstyle{=\ \, 1 + 28 \,l_1 + 294 \,l_1^2 + 1372 \,l_1^3 + 2401 \,l_1^4 + 33 \,l_2 + 693 \,l_1 l_2 + 
 4851 \,l_1^2 l_2 + 11319 \,l_1^3 l_2 + 407 \,l_2^2 + 5698 \,l_1 l_2^2 +}\\
 &&\quad \scriptstyle{ +19943 \,l_1^2 l_2^2 + 2223 \,l_2^3 + 15561 \,l_1 l_2^3 + 4536 \,l_2^4 + 34 \,l_3 + 
 714 \,l_1 l_3 + 4998 \,l_1^2 l_3 + 11662 \,l_1^3 l_3 + 838 \,l_2 l_3 + }\\
 &&\quad \scriptstyle{+11732 l_1 l_2 l_3 + 41062 l_1^2 l_2 l_3 + 6860 l_2^2 l_3 + 
 48020 l_1 l_2^2 l_3 + 18648 l_2^3 l_3 + 431 l_3^2 + 6034 l_1 l_3^2 + }\\
 &&\quad \scriptstyle{+21119 l_1^2 l_3^2 + 7051 l_2 l_3^2 + 49357 l_1 l_2 l_3^2 + 
 28728 l_2^2 l_3^2 + 2414 l_3^3 + 16898 l_1 l_3^3 + 19656 l_2 l_3^3 + }\\
 &&\quad \scriptstyle{+5040 l_3^4 + 36 l_4 + 756 l_1 l_4 + 5292 l_1^2 l_4 + 12348 l_1^3 l_4 + 
 887 l_2 l_4 + 12418 l_1 l_2 l_4 + 43463 l_1^2 l_2 l_4 + 7258 l_2^2 l_4 + }\\
&&\quad \scriptstyle{+50806 l_1 l_2^2 l_4 + 19719 l_2^3 l_4 + 912 l_3 l_4 + 12768 l_1 l_3 l_4 + 
 44688 l_1^2 l_3 l_4 + 14914 l_2 l_3 l_4 + 104398 l_1 l_2 l_3 l_4 + }\\
 &&\quad \scriptstyle{+60732 l_2^2 l_3 l_4 + 7656 l_3^2 l_4 + 53592 l_1 l_3^2 l_4 + 
 62307 l_2 l_3^2 l_4 + 21294 l_3^3 l_4 + 482 l_4^2 + 6748 l_1 l_4^2 +}\\ 
 &&\quad \scriptstyle{+23618 l_1^2 l_4^2 + 7879 l_2 l_4^2 + 55153 l_1 l_2 l_4^2 + 
 32067 l_2^2 l_4^2 + 8086 l_3 l_4^2 + 56602 l_1 l_3 l_4^2 + }\\
 &&\quad \scriptstyle{+65772 l_2 l_3 l_4^2 + 33705 l_3^2 l_4^2 + 2844 l_4^3 + 19908 l_1 l_4^3 + 
 23121 l_2 l_4^3 + 23688 l_3 l_4^3 + 6237 l_4^4 + 37 l_5 + 777 l_1 l_5 + }\\
 &&\quad \scriptstyle{+5439 l_1^2 l_5 + 12691 l_1^3 l_5 + 912 l_2 l_5 + 12768 l_1 l_2 l_5 + 
 44688 l_1^2 l_2 l_5 + 7465 l_2^2 l_5 + 52255 l_1 l_2^2 l_5 + 20286 l_2^3 l_5 + }\\
 &&\quad \scriptstyle{+938 l_3 l_5 + 13132 l_1 l_3 l_5 + 45962 l_1^2 l_3 l_5 + 15342 l_2 l_3 l_5 + 
 107394 l_1 l_2 l_3 l_5 + 62482 l_2^2 l_3 l_5 + 7877 l_3^2 l_5 + }\\
 &&\quad \scriptstyle{+55139 l_1 l_3^2 l_5 + 64106 l_2 l_3^2 l_5 + 21910 l_3^3 l_5 + 991 l_4 l_5 + 
 13874 l_1 l_4 l_5 + 48559 l_1^2 l_4 l_5 + 16204 l_2 l_4 l_5 + }\\
 &&\quad \scriptstyle{+113428 l_1 l_2 l_4 l_5 + 65961 l_2^2 l_4 l_5 + 16632 l_3 l_4 l_5 + 
 116424 l_1 l_3 l_4 l_5 + 135296 l_2 l_3 l_4 l_5 + 69335 l_3^2 l_4 l_5 + }\\
 &&\quad \scriptstyle{+8771 l_4^2 l_5 + 61397 l_1 l_4^2 l_5 + 71316 l_2 l_4^2 l_5 + 
 73066 l_3 l_4^2 l_5 + 25641 l_4^3 l_5 + 509 l_5^2 + 7126 l_1 l_5^2 + }\\
 &&\quad \scriptstyle{+24941 l_1^2 l_5^2 + 8325 l_2 l_5^2 + 58275 l_1 l_2 l_5^2 + 
 33894 l_2^2 l_5^2 + 8546 l_3 l_5^2 + 59822 l_1 l_3 l_5^2 + }\\
 &&\quad \scriptstyle{+69524 l_2 l_3 l_5^2 + 35630 l_3^2 l_5^2 + 9010 l_4 l_5^2 + 
 63070 l_1 l_4 l_5^2 + 73269 l_2 l_4 l_5^2 + 75068 l_3 l_4 l_5^2 +}\\ 
 &&\quad \scriptstyle{+39501 l_4^2 l_5^2 + 3083 l_5^3 + 21581 l_1 l_5^3 + 25074 l_2 l_5^3 + 
 25690 l_3 l_5^3 + 27027 l_4 l_5^3 + 6930 l_5^4.}
\end{eqnarray*}
Therefore we find that the number of lattice paths from $(0,0)$ to $(4n,3n)$, under the line $y=\frac{3}{4}x$, takes form
\begin{align}
\begin{split}
\sum_{l_1+\cdots+l_5=n}{b_{l_1,\ldots,l_5}(1)} &= \sum_{l_1+\cdots+l_5=n} A_{(3,4)}(l_1,l_2,l_3,l_4,l_5)\times \\
& \times\frac{1}{7 l_1+7 l_2+7 l_3+7l_4 +7 l_5+1}{7 l_1+7 l_2+7 l_3+7l_4 +7 l_5+1\choose l_1} \times\\
& \times\frac{1}{7 l_1+9 l_2+8 l_3+9l_4 +9 l_5+1}{7 l_1+9 l_2+8 l_3+9l_4 +9 l_5+1\choose l_2} \times \\
& \times\frac{1}{7 l_1+8 l_2+9 l_3+9l_4 +10 l_5+1}{7 l_1+8 l_2+9 l_3+9 l_4 +10 l_5+1\choose l_3} \times\\
& \times\frac{1}{7 l_1+9 l_2+9 l_3+11 l_4 +11 l_5+1}{7 l_1+9 l_2+9 l_3+11 l_4 +11 l_5+1\choose l_4} \times \\
& \times\frac{1}{7 l_1+9 l_2+10 l_3+11 l_4 +13 l_5+1}{7 l_1+9 l_2+10 l_3+ 11 l_4 +13  l_5+1\choose l_5}.
\end{split}
\end{align}
This expression does not seem to have been known before.

For $(3,4)$ torus knot we can also find an identity analogous to (\ref{ex-bridge-23}). First, for the quiver $C^{(3,4)}$ we find that $A_{(3,4)}^{\max}$, i.e. the homogeneous part of $A_{(3,4)}$, defined in (\ref{forAmax1}), takes form 
\begin{eqnarray*}
\scriptstyle{A_{(3,4)}^{\max}}(l_1,l_2,l_3,l_4,l_5)&=& \scriptstyle{2401 l_1^4 + 11319 l_1^3 l_2 + 19943 l_1^2 l_2^2 + 15561 l_1 l_2^3 + 
 536 l_2^4 + 11662 l_1^3 l_3 + 1062 l_1^2 l_2 l_3 + 8020 l_1 l_2^2 l_3 +} \\
&& 
 \scriptstyle{+18648 l_2^3 l_3 + 1119 l_1^2 l_3^2 + 9357 l_1 l_2 l_3^2 + 8728 l_2^2 l_3^2 + 
 16898 l_1 l_3^3 + 19656 l_2 l_3^3 + 5040 l_3^4 + 12348 l_1^3 l_4 + }\\
 &&
 \scriptstyle{+3463 l_1^2 l_2 l_4 + 50806 l_1 l_2^2 l_4 + 19719 l_2^3 l_4 + 
 4688 l_1^2 l_3 l_4 + 104398 l_1 l_2 l_3 l_4 + 60732 l_2^2 l_3 l_4 + }\\
 && \scriptstyle{+53592 l_1 l_3^2 l_4 + 62307 l_2 l_3^2 l_4 + 1294 l_3^3 l_4 + 618 l_1^2 l_4^2 + 
 55153 l_1 l_2 l_4^2 + 067 l_2^2 l_4^2 + 56602 l_1 l_3 l_4^2 + }\\
 && \scriptstyle{+65772 l_2 l_3 l_4^2 + 705 l_3^2 l_4^2 + 19908 l_1 l_4^3 + 121 l_2 l_4^3 + 
 688 l_3 l_4^3 + 6237 l_4^4 + 12691 l_1^3 l_5 + 4688 l_1^2 l_2 l_5 +} \\
 && \scriptstyle{+52255 l_1 l_2^2 l_5 + 20286 l_2^3 l_5 + 5962 l_1^2 l_3 l_5 + 
 107394 l_1 l_2 l_3 l_5 + 62482 l_2^2 l_3 l_5 + 55139 l_1 l_3^2 l_5 +}\\ 
 && \scriptstyle{+64106 l_2 l_3^2 l_5 + 1910 l_3^3 l_5 + 8559 l_1^2 l_4 l_5 + 
 113428 l_1 l_2 l_4 l_5 + 65961 l_2^2 l_4 l_5 + 116424 l_1 l_3 l_4 l_5 +} \\
 && \scriptstyle{+135296 l_2 l_3 l_4 l_5 + 69335 l_3^2 l_4 l_5 + 61397 l_1 l_4^2 l_5 + 
 71316 l_2 l_4^2 l_5 + 73066 l_3 l_4^2 l_5 + 5641 l_4^3 l_5 + 941 l_1^2 l_5^2 + }\\
 && \scriptstyle{+58275 l_1 l_2 l_5^2 + 894 l_2^2 l_5^2 + 59822 l_1 l_3 l_5^2 + 
 69524 l_2 l_3 l_5^2 + 5630 l_3^2 l_5^2 + 63070 l_1 l_4 l_5^2 +} \\
 && \scriptstyle{+73269 l_2 l_4 l_5^2 + 75068 l_3 l_4 l_5^2 + 9501 l_4^2 l_5^2 + 
 1581 l_1 l_5^3 + 5074 l_2 l_5^3 + 5690 l_3 l_5^3 + 7027 l_4 l_5^3 + 6930 l_5^4.}
\end{eqnarray*}
Again by relating the number of excursion and bridges via (\ref{bridges}) for $(4,-3)$ path, we find an equality
\begin{eqnarray*}
\frac{1}{7n}{7n \choose 3n}& =\sum\limits_{l_1+\ldots+l_5=n}  A_{(3,4)}^{\max}(l_1,l_2,l_3,l_4,l_5)\,\,\,\frac{1}{7 l_1+7 l_2+7 l_3+7l_4 +7 l_5}{7 l_1+7 l_2+7 l_3+7l_4 +7 l_5\choose l_1} \times \\
&  \frac{1}{7 l_1+9 l_2+8 l_3+9l_4 +9 l_5}{7 l_1+9 l_2+8 l_3+9l_4 +9 l_5\choose l_2}\frac{1}{7 l_1+8 l_2+9 l_3+9l_4 +10 l_5}{7 l_1+8 l_2+9 l_3+9 l_4 +10 l_5\choose l_3} \times \\
&  \frac{1}{7 l_1+9 l_2+9 l_3+11 l_4 +11 l_5}{7 l_1+9 l_2+9 l_3+11 l_4 +11 l_5\choose l_4}\frac{1}{7 l_1+9 l_2+10 l_3+11 l_4 +13 l_5}{7 l_1+9 l_2+10 l_3+ 11 l_4 +13  l_5\choose l_5}.
\end{eqnarray*}


\subsection{Reconstructing quivers from the Bizley formula}

In previous examples we showed that indeed generating functions of lattice paths are reproduced by generating functions of colored torus knot polynomials, or appropriate quiver generating functions. Now we illustrate that one can in fact reconstruct the quiver from the knowledge of the generating function. This should indeed be possible: the classical quiver generating function is determined by a finite set of parameters, i.e. entries of a matrix $C$ that encodes the quiver, and coefficients of this function are of the form (\ref{b1-glavna}). Therefore, once we know (from some other source) sufficient number of coefficients of this generating function, we should be able to reconstruct the form of the matrix $C$. This is valid even when diagonal invariants are considered -- in this case one should simply compare more coefficients of the generating function.

Let us illustrate this procedure in the example of paths under the line of the slope $\frac{2}{3}$. Assume that the generating function of such paths is encoded in a quiver with two vertices, determined by a matrix (\ref{2vert}) with some unknown entries $\alpha$, $\beta$ and $\gamma$, and consider the quiver generating function $y(x)$ with identified generating parameters $x=x_1=x_2$. Instead of the generating function itself it is convenient to write down its logarithm (\ref{log-y-x1x2}), and its expansion to the fourth order takes form
\begin{align}
\begin{split}
\log y(x) &= 2 x+(-1+\alpha+2 \beta+\gamma) x^2+\\
&+\frac{1}{6} (4-9 \alpha+9 \alpha^2-18 \beta+18 \alpha \beta+18 \beta^2-9 \gamma+18 \beta \gamma+9 \gamma^2) x^3+\\
&+\frac{1}{6} (-3+11 \alpha-24 \alpha^2+16 \alpha^3+22 \beta-48 \alpha \beta+36 \alpha^2 \beta-48 \beta^2+48 \alpha \beta^2+32 \beta^3+\\
&\qquad +11 \gamma-48 \beta \gamma+24 \alpha \beta \gamma+48 \beta^2 \gamma-24 \gamma^2+36 \beta \gamma^2+16 \gamma^3) x^4+\ldots
\end{split}
\end{align}
On the other hand, we suspect that this generating function should count lattice paths under the line of the slope $\frac{2}{3}$, which are given by the Bizley formula (\ref{y-Bizley}) or (\ref{paths-23}), whose logarithm for such paths takes form
\be
\log y(x) = \sum_{n=0}^{\infty} \frac{1}{5n}\binom{5n}{2n} x^n = 2x+21x^2 + \frac{1001}{3} x^3 + \frac{12597}{2}x^4 +\ldots
\ee
Comparing coefficients at $x^2$, $x^3$ and $x^4$ in the above two expansions gives a set of three equations, which determine three entries of the quiver matrix in either of two equivalent forms
\be
(\alpha, \beta,\gamma) = (5,5,7)\qquad\textrm{or} \qquad (\alpha, \beta,\gamma) = (7,5,5).
\ee
In this way we indeed reconstruct the quiver for the trefoil knot (\ref{CT23}). 

In principle, with enough computational power, from the Bizley formula one could reconstruct a quiver that encodes path counting for any slope $r/s$. At the same time, such quivers would encode colored extremal ($q$-dependent) polynomials for arbitrary torus knots, and also $q$-weighted path numbers. It is amusing that, at least in principle, colored extremal HOMFLY-PT polynomials for all torus knots, and $q$-weighted path numbers, are encoded in a relatively simple Bizley formula (\ref{y-Bizley}).


\subsection{Generalized Bizley formula}

For completeness let us derive a version of the Bizley formula in an arbitrary framing, once this formula is interpreted as the generating function of torus knot invariants. A change of framing of a knot by $f$ has the following effect on the generating function
\be \label{gen_func_framing}
y_K^{(f)}(x) = y_K(x (y_K^{(f)}(x))^f).
\ee
Once we know the function $y_K(x)$ we can use the Faa di Bruno formula to express coefficients of $y_K^{(f)}(x)$ through coefficients of $y_K(x)$. Formula like this, in the special case of relating number of factor-free words with a number of all lattice paths, appeared already in~\cite{2016arXiv160602183B}. The generating function $y_K(x)$ for the $(r,s)$ torus knot with framing $rs$ is given by the Bizley formula (\ref{y-Bizley}). 
\begin{proposition}
Generalization of the Bizley formula to framed invariants reads
\be
y_K^{(f)} = \exp\Big(\sum_{k=1}^{\infty} c_n^{(f)} x^n \Big) = \sum_{n=0}^{\infty} b_n^{(f)} x^n,  \label{y-Bizley-f}
\ee
with
\begin{align}
\begin{split}
c_n^{(f)} &= \frac{1}{n f\cdot n!}   \sum_{k=1}^n B_{n,k}(n f g_1 1!, nf g_2 2!, nfg_3 3!, \ldots),  \\
b_n^{(f)} &= \frac{1}{nf}\frac{1}{n!} \sum_{k=1}^n \frac{(1+nf)!}{(1+nf-k)!} B_{n,k}(b_1 1!, b_2 2!, \dots b_{n-k+1}(n-k+1)!). \label{y-Bizley-f-2}
\end{split}
\end{align}
Here the framing $f$ is defined with respect to the framing~$rs$ and $B_{n,k}$ are partial Bell polynomials 
\be
B_{n,k}(x_1, x_2, \ldots, x_{n-k+1}) = \sum_{\{p_j\}} \frac{n!}{p_1! p_2! \cdots} \left(\frac{x_1}{1!}\right)^{p_1} \left(\frac{x_2}{2!}\right)^{p_2} \cdots \left(\frac{x_{n-k+1}}{(n-k+1)!}\right)^{p_{n-k+1}},
\ee
where the summation extends over all sets of numbers $\{p_j\}$ such that
\be
\sum_{j} p_j = k,\qquad \sum_j j p_j = n. 
\ee
For $f=0$ we get the original Bizley formula (\ref{y-Bizley})
\be
c_n^{(0)} = g_n = \frac{1}{(r+s)n}\binom{(r+s)n}{rn}.
\ee
\end{proposition}

Let us prove the formula for $b_n^{(f)}$ in (\ref{y-Bizley-f-2}). Lagrange inversion theorem applied to~\eqref{gen_func_framing} results in the following relation
\be
[x^n] y_K^{(f)}(x) = \frac{1}{1+nf} [x^n](y_K(x))^{1+nf}, \label{lagr_inverse}
\ee
where $[x^n]y(x)$ denotes a coefficient of $x^n$ in the expansion of $f(x)$. Denote $[x^n]y_K(x) = b_n$. A power of the generating function can be computed using the Faa di Bruno formula for a derivative of a composite function
\be
\frac{\partial^n}{\partial x^n} h(x) = \sum_{k=1}^n f^{(k)}(y_0(x)) B_{n,k}(y_0^{(1)}(x),y_0^{(2)}(x), \dots, y_0^{(n-k+1)}(x)).
\ee
Define
\be
f(x) = x^{1+nf}, \qquad h(x) = f(y_K(x)).
\ee
Then
\be
[x^n](y_K(x))^{1+nf} = \frac{1}{n!}\frac{\partial^n}{\partial x^n} h(x)|_{x=0}.
\ee
Evaluating the $k$-th derivative of $f$ gives
\be
f^{(k)}(x) = \begin{cases}
  \frac{(1+nf)!)}{(1+nf - k)!} x^{1+nf-k}, \qquad &k \leq 1+nf\\
  0 \qquad &k > 1+nf
\end{cases}
\ee
The condition $k\leq 1+nf$ is always fulfilled (for positive $f$) because $k \leq n$. Then, using $y_0(0)=1$, we find
\be
[x^n](y_K(x))^{1+nf}  = \frac{1}{n!} \sum_{k=1}^n \frac{(1+nf)!}{(1+nf-k)!} B_{n,k}(b_1 1!, b_2 2!, \dots b_{n-k+1}(n-k+1)!).
\ee
Finally using the relation~\eqref{lagr_inverse} we obtain (\ref{y-Bizley-f-2}).

As an illustration we explicitly list coefficients $b_n(f)$ for the trefoil in table \ref{tab-biz}. In this case the zero framing reproduces coefficients in (\ref{paths-23}), while results for $f=-5$ correspond to the generating function of factor-free words (see the next subsection~\ref{ssec-Apoly}).

\begin{table}[h!]
  \centering
\begin{tabular}{|c | l|}
\hline
  $f$ & $b_n^{(f)}$ \\ \hline
  $-5$ & 1, 2, 3, 7, 19, 56, 174, 561, \ldots \\
  $-4$ & 1, 2, 7, 33, 181, 1083, 6854, 45111, \ldots \\
  $-3$ & 1, 2, 11, 83, 727, 6940, 70058, 735502, \ldots \\
  $-2$ & 1, 2, 15, 157, 1913, 25427, 357546, 5229980, \ldots \\
  $-1$ & 1, 2, 19, 255, 3995, 68344, 1237526, 23316295, \ldots \\
  $0$ & 1, 2, 23, 377, 7229, 151491, 3361598, 77635093, \ldots\\
  $1$ & 1, 2, 27, 523, 11871, 294668, 7747698, 212054604, \ldots\\
  $2$ & 1, 2, 31, 693, 18177, 521675, 15863042, 502196626, \ldots\\
\hline
\end{tabular}
\caption{Coefficients of framed Bizley generating function.}  \label{tab-biz}
\end{table}

Moreover, results in this table agree with the results of section~\ref{sec31} for coefficients $b_{i,j}(1)$ of a quiver
\ea{
C=\left[
\begin{array}{cc}
f+7 & f+5\\
f+5 & f+5
\end{array}
\right],
}
upon the identification $b_n^{(f)} = \sum_{i+j=n} b_{i,j}(1)$.


\subsection{Algebraic equations and extremal A-polynomials}   \label{ssec-Apoly}

Finally we illustrate, and prove in several cases, the relation between A-polynomials and equations satisfied by generating functions of paths. As we explained in section \ref{ssec-paths}, generating functions (\ref{Paths-x}) of lattice paths under the line of the slope $\frac{r}{s}$ satisfy algebraic equations, which can be determined e.g. from the Duchon grammar. Proposition \ref{prop2} implies that these equations should be the same as extremal A-polynomial equations for $(r,s)$ torus knots in framing $rs$. Extremal A-polynomials can be computed by the saddle point method from the knowledge of colored extremal invariants, or equivalently from the knowledge of the corresponding quiver and the formula (\ref{PxC}). For various knots such computations have been conducted in \cite{Garoufalidis:2015ewa}. Examples of such algebraic equations for several knots, in framing $rs$, are given in table \ref{tab-A}. It is straightforward to check to arbitrarily high order, that generating functions of lattice paths, given by the Bizley formula (\ref{y-Bizley}) or our result (\ref{yPx-prop}), satisfy these algebraic equations.

\begin{table}[h]
\begin{small}
\be
\begin{array}{|c|c|}
\hline 
\textrm{\bf Paths / torus knot} & A(x,y)   \nonumber \\
\hline 
\hline
(2,3) & 1 -y + x ( 2 y^5 - y^6 + y^7 ) + x^2 y^{10} \\
\hline
(2,5) &  \ 1-y + x (3 y^7 -  2 y^8 +2 y^9 - y^{10} + y^{11} ) + x^2 (3 y^{14} - y^{15} + 2y^{16}) + x^3 y^{21}  \ \\
\hline
(2,7) & 1 -y + x y^9 (4- 3 y + 3 y^2 - 2 y^3 + 2 y^4 - y^5 + y^6) +\\
    &  + x^2 y^{18}(6  - 3 y + 6 y^2 - 2 y^3 + 3 y^4)+ x^3 y^{27} (4- y + 3 y^2) + x^4 y^{36}  \\
\hline
(3,4) & 1-y+ x y^7 (5- 4 y + y^2 +3 y^3 - y^5 + y^6) + \\
 & + x^2 y^{14} (10 - 6 y + 3 y^2 + 5 y^3 - y^4 + y^5) +\\
 &  + x^3 y^{21} (10 - 4 y + 3 y^2 + y^3 - y^4) + x^4 y^{28} (5 - y +y^2 - y^3) + x^5 y^{35} \\
\hline
(3,5) & 1-y +  x y^8 (7- 6 y + y^2 +5 y^3 - 3 y^4 + 3 y^5 - y^7 + y^8) + \\
 & + x^2 y^{16} (21 - 15 y + 5 y^2 + 18 y^3 - 9 y^4 + 5 y^5 +3 y^6) +\\
 & + x^3 y^{24} (35 - 20 y + 10 y^2 + 22 y^3 - 9 y^4 + 2 y^6 - 2 y^7) +\\
 & + x^4 y^{32} ( 35 - 15 y + 10 y^2  + 8 y^3 - 3 y^4 - 3 y^5) +\\
 & + x^5 y^{40} (21 - 6 y + 5 y^2- 3 y^3- y^5  + y^6) +\\
 & + x^6 y^{48} (7- y + y^2 - 2 y^3) + x^7 y^{56} \\
\hline
\end{array}
\ee  
\end{small}
\caption{Algebraic equations and extremal (top row, left-handed) A-polynomials for $(r,s)$ torus knots in framing $rs$.} \label{tab-A}
\end{table}

Apart from checking that generating functions of lattice paths satisfy A-polynomial equations we can also rederive these equations, taking advantage of the Duchon grammar. This proves to all orders that generating functions of lattice paths and knot polynomials are equal.

Let us first summarize Duchon's formalism \cite{Duchon}, which reformulates the problem of counting lattice paths in terms of constructing and counting words obeying certain grammar. The words are created from an alphabet, which in the case of lattice paths under the $y=\frac{r}{s} x$ line consists of two letters $U = \{a,b\}$. Denote by $U^*$ the set of all words in the alphabet $U$. The letters correspond to two steps that a lattice path is made of. The type of the lattice path counting problem is encoded in the valuation of the letters. We define a valuation function $h$ on the alphabet with values in integers such that $h(a) = r$ and $h(b) = -s$. This definition extends additively to the set of words $U^*$, e.g. $h(aaba) = 3r - s$. The lattice path counting problem can be then made equivalent to the problem of counting words. The condition, that a path in the upper half plane picture reaches back but never crosses the horizontal axis can be formulated with the valuation function. For the path to reach back to the horizontal axis, the valuation function of the corresponding word must be $0$. For the path to never cross the horizontal axis, the valuation function of every left factor of the word cannot be negative. A left factor $w_L$ of a word $w$ is simply any left part of the word $w$. The set of words obeying this constraint is denoted by $D_{r/s} \subset U^* $. 

Among words in $D_{r/s}$ there are special ones that cannot be generated from simpler words. For example, for $r=3$ and $s=2$ there are two words of length $5$
\be
ababb, \qquad aabbb.
\ee
Many words of length $10$ can be obtained by taking one of the word of length $5$ and using it as a template. Between any letters of this word we can insert any word of length $5$ to obtain a word of length $10$. For example taking $ababb$ as a template we can get $\mathbf{aba}aabbb\mathbf{bb}$ by inserting $aabbb$ between the third and the fourth letter. However there are words of length $10$ which cannot be obtained in this way, for example $aaabbabbbb$. Such words are called factor-free words. Duchon showed how to generate (and thus count) all factor-free words, and how to obtain a generating function of all words in $D_{r/s}$ from the generating function of factor-free words. 

Let us denote the generating function of factor-free words by $y_{fP}(x)$. Then the generating function $y_P(x)$ of all the words is given by
\be
y_P(x) = y_{fP}\big(x (y_P(x))^{r+s}\big),
\ee
This relation is equivalent to a change framing of the generating function by $r+s$. As an immediate consequence we obtain that the factor free words also find their place in the knots-paths correspondance and simply correspond to $(r,s)$ torus knots framed by $-r  s + r+s$. 

We describe now the construction of equations for $y_{fP}(x)$. Duchon showed that factor free words can be generated from the following grammar
\be
\tilde{D} = \epsilon + \sum_k \tilde{L}_k \tilde{R}_k,\qquad
\tilde{L}_i = \delta_{i,r} a + \sum_k \tilde{L}_k \tilde{R}_{k-i},\qquad
\tilde{R}_j = \delta_{j,s} b + \sum_k \tilde{L}_k \tilde{R}_{j+k},
\ee
with indices in the range $1 \geq i \geq r$, $1 \geq j \geq s$, and $\tilde{L}_i = \tilde{R}_j = 0$ for indices beyond this range. Here $\epsilon$ denotes an empty word   and in general the letters $a$ and $b$ do not commute. To construct short words it is enough to solve the equations iteratively. In the classical case, where we are interested in counting paths, letters $a$ and $b$ commute. Moreover each path must consists of $k s$ $a$ steps and $k r$ $b$ steps, so that the valuation of the whole path is $k sr - ksr =0$. Therefore the relevant variable is $x = a^s b^r$. Eliminating auxillary sets $\tilde{L}_i$ and $\tilde{R}_j$ we obtain a polynomial equation for $\tilde{D}(x)$.

As an example consider paths under the $y=\frac{3}{2} x$ line, corresponding to the trefoil knot. This is the case solved explicitly by Duchon. The set of equations takes form
\begin{align}
\begin{split}
 \tilde{D} &= \epsilon + \tilde{L}_1 \tilde{R}_1 + \tilde{L}_2 \tilde{R}_2, \qquad \tilde{L}_1 = \tilde{L}_2 \tilde{R}_1 + \tilde{L}_3 \tilde{R}_2,\\
  \tilde{L}_2 &= \tilde{L}_3 \tilde{R}_1, \qquad \tilde{L}_3 = a, \qquad\tilde{R}_1 = \tilde{L}_1 \tilde{R}_2, \qquad \tilde{R}_2 = b.
\end{split}
\end{align}
We eliminate $\tilde{L}_i$ and $\tilde{R}_2$ and parametrize $\tilde{R}_1 = a U b^2$ to find a set of two equations
\begin{align}
\begin{split}
\tilde{D} &= \epsilon + a^2 U b^3 + aba U b^2 + a^2 U b^2 a U b^2 a U b^2,\\
U &= \epsilon + a U b^2 a U b.
\end{split}
\end{align}
For the word counting problem we consider commuting $a$ and $b$ and introduce $x=a^2 b^3$, so that we obtain
\begin{align}
\begin{split}
y_{fP}(x) &= 1 + 2x U(x) + x^2 U^3(x) ,\\
U(x) &= 1 +  x^2 U(x).
\end{split}
\end{align}
On one hand $U(x)$ is the generating function of the Catalan numbers and using twice the equation for $U(x)$ in the equation for $y_{fP}(x)$ we get
\be
y_{fP}(x) = (1 + x)U(x) = \sum_{k=0}^{\infty} (C_k + C_{k+1})x^k.
\ee
On the other hand we can eliminate $U(x)$ to find the equation $A_f(x, y_{fP}) = 0$ which the generating function of factor-free words satisfies
\be
A_f(x,y_{fP}) = (1+x)^2 - y_{fP} - xy_{fP} + xy_{fP}^2.
\ee
Changing the framing by $r+s=5$ gives then the algebraic equation 
\be
A(x,y) = 1 -y + x ( 2 y^5 - y^6 + y^7 ) + x^2 y^{10},
\ee
which indeed reproduces the A-polynomial equation for $(2,3)$ torus knot given in table \ref{tab-A}.

Similar computations for paths under the $y=\frac{2}{5} x$ line, or equivalently for the $(2,5)$ torus knot, lead to the following set of equations 
\begin{align}
\begin{split}
  y_{fP}(x) &= 1 + 3 x U(x) + 4 x^2 U^3(x) + x^3 U^5(x),\\
  U(x) &= 1 + 3 x U^2(x) + x^2 U^4(x),
\end{split}
\end{align}
with $U$ defined this time through $\tilde{R}_1 = a U b^3$ and $x = a^2 b^5$. Eliminating $U$ we find the algebraic equation satisfied by the generating function of factor-free words
\be
A_f(x,y_{fP}) = 1 - y_{fP} + x(3 -2y_{fP} +2y_{fP}^2 - y_{fP}^3 +y_{fP}^4) + x^2(3 - y_{fP} + 2y_{fP}^2) + x^3.
\ee
Changing the framing by $r+s=7$ produces then the A-polynomial equation for $(2,5)$ torus knot, which is given in table \ref{tab-A}.


\section{Knot polynomials, quivers and path counting for $(3,s)$ torus knots}     \label{sec-torus3p}

In this section we identify quivers and extremal colored HOMFLY-PT polynomials for a class of $(3,s)$ torus knots, for $s=3p+1$ or $s=3p+2$. This is quite a non-trivial class of examples, which nicely illustrates the power of the knots-quivers correspondence, as well as the relation of torus knot invariants to the counting of lattice paths. Indeed, it is straightforward to verify that expressions for colored polynomials for torus knots given below, in appropriate framing, agree with generating functions of lattice paths, as we explained earlier. 


Our strategy is similar as in other examples of knots-quivers correspondence: we consider  extremal colored HOMFLY-PT polynomials for first few symmetric colors, and -- also based on the knowldege of homological degrees encoded in the uncolored extremal superpolynomial -- we identify a matrix encoding the corresponding quiver uniquely. In particular we find various regularities, which enable to analyze at once the whole classes of $(3,3p+1)$ and $(3,3p+2)$ torus knots and reveal their recursive structure.

\subsection{$(3,3p+2)$ torus knots}

The constraints that we found upon the analysis of several first representations imply, that the quiver matrix for $(3,3p+2)$ knots has the following structure
\begin{align}
  C^{(3,3p+2)} = \begin{bmatrix}
    C^{(2, 6p+3;-3)} & B_p \\
    B_p & C^{(3, 3p-1;+5)}
  \end{bmatrix}
\end{align}
Here $C^{(r,s;f)}$ denotes a quiver matrix for $(r,s)$ torus knot with an additional framing, $f$ with respect to the convention explained at the end of the section~\ref{ssec-knots-quivers}. For example, for $p=1$, the bottom right block is $C^{(3,2+5)}$ which is the trefoil quiver~\eqref{CT23} with all entries shifted by $+5$.  The off-diagonal rectangular block $B$ carries further information about the recursive structure and it is organized in the following way 
\ea {
  B_p = \begin{bmatrix}
    B_{p,p} & B_{p,p-1} & \dots & B_{p,1} 
    \end{bmatrix},
}
where each $B_{p,k}$ (for $1 \leq k \leq p$) is a matrix of size
\ea {
  (3p+2) \times (3k-1).
}
The block $B_{p,k}$ is composed of two groups of rows
\ea {
  B_{p,k} = \begin{bmatrix}
    X_{p,k} \\
    Y_{p,k}
  \end{bmatrix}
}
of sizes
\be
X_{p,k}: \quad (3(p-k+1)) \times (3k-1), \qquad \qquad Y_{p,k}: \quad (3k-1) \times (3k-1).
\ee
The  matrices $X$ and $Y$ have the following structure
\bea
X_{p,k} &=&  \begin{bmatrix}
6p+3k+4 & 6p+3k+2 & 6p+3k & \cdots & 6p-3k+8\\
6p+3k+3 & 6p+3k+1 & 6p+3k-1 & \cdots & 6p-3k+7\\
6p+3k+2 & 6p+3k & 6p+3k-2 & \cdots & 6p-3k+6\\
\vdots  & \vdots & \vdots & \ddots & \vdots\\
3p+6k+2 & 3p+6k & 3p+6k-2 & \cdots & 3p+6
  \end{bmatrix}, \\\nonumber\\\nonumber\\
  Y_{p,k} &=& \begin{bmatrix}
3p+6k+1 & 3p+6k-1 & 3p+6k-3 & \cdots & 3p+5\\
3p+6k & 3p+6k-1 & 3p+6k-3 & \cdots & 3p+5\\
3p+6k-2 & 3p+6k-2 & 3p+6k-3 & \cdots & 3p+5\\
\vdots  & \vdots & \vdots & \ddots & \vdots\\
3p+6 & 3p+6 & 3p+6 & \cdots & 3p+5
    \end{bmatrix}.
\eea

To illustrate the above structure, let us consider the first two knots in this series. The value $p=1$ corresponds to $(3,5)$ torus knot. In this case we only have $B_{1,1}$ matrix, which is built out of $X_{1,1}$ and $Y_{1,1}$, and takes form
\ea {
  B_{1,1} = \begin{bmatrix}
    13 & 11 \\
    12 & 10 \\
    11 & 9 \\
    10 & 8 \\
    9 & 8
  \end{bmatrix}.
}
The full quiver matrix, with the block structure highlighted, takes form
\ea {
  C^{(3,5)} =  \begin{bmatrix}
    C^{(2,9;-3)}  & B_1 \\
    B_1 & C^{(3,2;+5)}
    \end{bmatrix} = \begin{bmatrix}
 16 & 14 & 12 & 10 & 8 & 13 & 11 \\
 14 & 14 & 12 & 10 & 8 & 12 & 10 \\
 12 & 12 & 12 & 10 & 8 & 11 & 9 \\
 10 & 10 & 10 & 10 & 8 & 10 & 8 \\
 8 & 8 & 8 & 8 & 8 & 9 & 8 \\
 13 & 12 & 11 & 10 & 9 & 12 & 10 \\
 11 & 10 & 9 & 8 & 8 & 10 & 10 \\
\end{bmatrix}
}
The second example, for $p=2$, corresponds to $(3,8)$ torus knot. In this case we introduce two matrices $B_{2,1}$ and $B_{2,2}$, which are given explicitly by
\ea {
  B_{2,2} = \begin{bmatrix}
 22 & 20 & 18 & 16 & 14 \\
 21 & 19 & 17 & 15 & 13 \\
 20 & 18 & 16 & 14 & 12 \\
 19 & 17 & 15 & 13 & 11 \\
 18 & 17 & 15 & 13 & 11 \\
 16 & 16 & 15 & 13 & 11 \\
 14 & 14 & 14 & 13 & 11 \\
 12 & 12 & 12 & 12 & 11 \\
  \end{bmatrix} \qquad \qquad
  B_{2,1} = \begin{bmatrix}
 19 & 17 \\
 18 & 16 \\
 17 & 15 \\
 16 & 14 \\
 15 & 13 \\
 14 & 12 \\
 13 & 11 \\
 12 & 11 \\
  \end{bmatrix}
}
The full quiver matrix, with the block structure highlighted, reads
\ea {
  C^{(3,8)}  &= \begin{bmatrix}
    C^{(2,15;-3)}  & B_2 \\
    B_2 & C^{(3,5;+5)}
  \end{bmatrix} = 
\begin{bmatrix}
 25 & 23 & 21 & 19 & 17 & 15 & 13 & 11 & 22 & 20 & 18 & 16 & 14 & 19 & 17 \\
 23 & 23 & 21 & 19 & 17 & 15 & 13 & 11 & 21 & 19 & 17 & 15 & 13 & 18 & 16 \\
 21 & 21 & 21 & 19 & 17 & 15 & 13 & 11 & 20 & 18 & 16 & 14 & 12 & 17 & 15 \\
 19 & 19 & 19 & 19 & 17 & 15 & 13 & 11 & 19 & 17 & 15 & 13 & 11 & 16 & 14 \\
 17 & 17 & 17 & 17 & 17 & 15 & 13 & 11 & 18 & 17 & 15 & 13 & 11 & 15 & 13 \\
 15 & 15 & 15 & 15 & 15 & 15 & 13 & 11 & 16 & 16 & 15 & 13 & 11 & 14 & 12 \\
 13 & 13 & 13 & 13 & 13 & 13 & 13 & 11 & 14 & 14 & 14 & 13 & 11 & 13 & 11 \\
 11 & 11 & 11 & 11 & 11 & 11 & 11 & 11 & 12 & 12 & 12 & 12 & 11 & 12 & 11 \\
 22 & 21 & 20 & 19 & 18 & 16 & 14 & 12 & 21 & 19 & 17 & 15 & 13 & 18 & 16 \\
 20 & 19 & 18 & 17 & 17 & 16 & 14 & 12 & 19 & 19 & 17 & 15 & 13 & 17 & 15 \\
 18 & 17 & 16 & 15 & 15 & 15 & 14 & 12 & 17 & 17 & 17 & 15 & 13 & 16 & 14 \\
 16 & 15 & 14 & 13 & 13 & 13 & 13 & 12 & 15 & 15 & 15 & 15 & 13 & 15 & 13 \\
 14 & 13 & 12 & 11 & 11 & 11 & 11 & 11 & 13 & 13 & 13 & 13 & 13 & 14 & 13 \\
 19 & 18 & 17 & 16 & 15 & 14 & 13 & 12 & 18 & 17 & 16 & 15 & 14 & 17 & 15 \\
 17 & 16 & 15 & 14 & 13 & 12 & 11 & 11 & 16 & 15 & 14 & 13 & 13 & 15 & 15
\end{bmatrix}
}

We compute now the generating functions of lattice paths using the relation to knots and quivers (\ref{yPx-prop}). In the Table~\ref{tab-3p2} we present such generating functions for the first 5 knots of the series $(3,3p+2)$. These numbers agree with numbers of lattice paths that follow from the Bizley formula~\eqref{y-Bizley}.

\begin{table}[h!]
  \centering
\begin{tabular}{|c | l|}
\hline
  \textrm{\bf Knot} & \qquad $b_n$ (numbers of lattice paths) \\ \hline
  (3,5) & 1, 7, 525, 58040, 7574994, 1084532963, 164734116407, \ldots \\
  (3,8) & 1, 15, 3504, 1220135, 502998985, 227731502703, 109447217699997, \ldots \\
  (3,11) & 1, 26, 13793, 10969231, 10342244094, 10714942416045, 11787169120183931,  \ldots \\
  (3,14) & 1, 40, 40356, 61246090, 110288829466, 218304920579248, \ldots \\
  (3,17) & 1, 57, 97584, 251886268, 771887463392, \ldots\\
\hline
\end{tabular}
\caption{Numbers of lattice paths under the $y=\frac{3}{3p+2}x$ line. They agree with the Bizley formula, and with coefficients of classical generating functions of invariants of $(3,3p+2)$ torus knot.}  \label{tab-3p2}
\end{table}

The other side of the knots--quivers--paths correspondence yields colored HOMLFY-PT polynomials. To this end we need to identify, using the uncolored (extremal) HOMFLY-PT homology, the variables $x_i$ as prescribed in eq.~\eqref{specialize}. The $t$ degrees are equal to the diagonal entries of the quiver matrix
\be
t_i = C_{ii}, \label{tdegrees}
\ee
whereas the $q$ degrees are an ordered union of sets $Q_n$ for $n=p,\dots, 0$ with
\be
Q_n = \{ -6n - 2, -6n +2, \dots 6n - 2, 6n +2 \}.
\ee
The ordering is such that the set of $q$ degrees starts with $Q_p$ and the other follow in the descending order. 
For example for the $(3,5)$ and $(3,8)$ knots the $q$ degrees are
\bea
&&\{-8, -4, 0, 4, 8, -2, 2 \},\\
&&\{-14, -10, -6, -2, 2, 6, 10, 14, -8, -4, 0, 4, 8, -2, 2, \}
\eea
To obtain the standard form of the HOMFLY-PT polynomial -- right-handed with the zero framing -- the quiver matrix has to be transformed as explained in the section~\ref{ssec-knots-quivers}. For example the $C^{(3,5)}$ quiver in this case becomes
\ea {
  C^{(3,5)} =  \begin{bmatrix}
 0 & 1 & 3 & 5 & 7 & 2 & 4 \\
 1 & 2 & 3 & 5 & 7 & 3 & 5 \\
 3 & 3 & 4 & 5 & 7 & 4 & 6 \\
 5 & 5 & 5 & 6 & 7 & 5 & 7 \\
 7 & 7 & 7 & 7 & 8 & 6 & 7 \\
 2 & 3 & 4 & 5 & 6 & 4 & 5 \\
 4 & 5 & 6 & 7 & 7 & 5 & 6 
\end{bmatrix}
}
The $t$ degrees defined in~\eqref{tdegrees} can be read off from the diagonal of this matrix. 

From quivers that we found above, extremal colored HOMLFY-PT polynomials for $(3,3p+2)$ torus knots can be determined using (\ref{PxC-ext}). Examples of such minimal polynomials for several (right-handed) knots (after adjusting the quiver so that it captures minimal invariants of right-handed knots), in the fundamental representation, are given in table \ref{tab-33p2-homly}. These results agree with known such polynomials in the fundamental representation -- however, we stress that from the quivers determined above we also immediately get formulas for knot polynomials in arbitrary symmetric representations, which have not been known before.

\begin{table}[h!]
  \begin{small}
    \be
    \begin{array}{|c|c|}
      \hline 
      \textrm{\bf Torus knot} & P_1^{-}(q)   \nonumber \\
      \hline 
      \hline
      (3,5) & q^{-8} + q^{-4} + q^{-2} + 1 + q^2 + q^4 + q^8 \\
      \hline
      (3,8) &  q^{-14} + q^{-10} + q^{-8} + q^{-6} + q^{-4} + 2 q^{-2} + 1 + 2q^{2} + q^4 + q^6 + q^8 + q^{10} + q^{14}  \ \\
      \hline
      (3,11) & q^{-20} + q^{-16} + q^{-14} + q^{-12} + q^{-10} + 2q^{-8} + q^{-6} + 2q^{-4} + 2q^{-2} + 2 + 2q^2 + 2q^4 \\
      & + q^6 + 2q^8 + q^{10} + q^{12} + q^{14} + q^{16} + q^{20}\\
      \hline
      (3,14) &  q^{-26} + q^{-22} + q^{-20} + q^{-18} + q^{-16} + 2q^{-14} + q^{-12} + 2 q^{-10} + 2 q^{-8} + 2 q^{-6} + 2q^{-4} \\
                              & + 3 q^{-2} + 2 + 3q^2 + 2q^4 + 2q^6 + 2q^8 + 2q^{10} + q^{12} + 2 q^{14} + q^{16} + q^{18} \\ &+ q^{20} + q^{22} + q^{26}\\
      \hline
      (3,17) & q^{-32} + q^{-28} + q^{-26} + q^{-24} + q^{-22} + 2 q^{-20} + q^{-18} + 2q^{-16} + 2q^{-14} + 2q^{12} \\
                              & + 2q^{-10} + 3q^{-8} + 2q^{-6} + 3q^{-4} + 3q^{-2} + 3 + 3 q^2 + 3q^4 + 2q^6 + 3q^8 + 2q^{10}\\
                              &+ 2q^{12} + 2q^{14} + 2q^{16} + q^{18} + 2q^{20} + q^{22} + q^{24} + q^{26} + q^{28} + q^{32}\\
      \hline
    \end{array}
    \ee
  \end{small}
  \caption{Minimal HOMFLY-PT polynomials for right-handed $(3,3p+2)$ torus knots in the trivial framing, in the fundamental representation.}   \label{tab-33p2-homly}
\end{table}

Finally, having found quivers for $(3,3p+2)$ torus knots, we can also determine $q$-weighted path numbers using (\ref{qPaths-quiver}). The first such numbers, i.e. $q$-numbers of paths between the origin and the point with coordinates $(3p+2,3)$, for several knots, are given in table \ref{tab-33p2-qpaths}. For $q=1$ these expressions reduce to unweighted path numbers given in given in table \ref{tab-3p2}.

\begin{table}[h!]
  \begin{small}
    \be
    \begin{array}{|c|l|c|}
      \hline
      \textrm{\bf Paths} & \qquad b_1(q) & \ b_1(1) \ \nonumber \\
      \hline
      \hline
      (3,5) & q^7 + 2q^9 + 2q^{11} + q^{13} + q^{15} & 7 \\
      \hline
      (3,8) &  q^{10}+2 q^{12}+3 q^{14}+3 q^{16}+2 q^{18}+2 q^{20}+q^{22}+q^{24} & 15\\
      \hline
      (3,11) & q^{13}+2 q^{15}+3 q^{17}+4 q^{19}+4 q^{21}+3 q^{23}+3 q^{25}+2 q^{27}+2 q^{29}+q^{31}+q^{33} & 26\\
      \hline
      (3,14) &  q^{16}+2 q^{18}+3 q^{20}+4 q^{22}+5 q^{24}+5 q^{26}+4 q^{28}+4 q^{30}+3 q^{32} &40\\
      & + 3 q^{34}+2 q^{36}+2 q^{38}+q^{40}+q^{42} &\\
      \hline
      (3,17) & q^{19}+2 q^{21}+3 q^{23}+4 q^{25}+5 q^{27}+6 q^{29}+6 q^{31}+5 q^{33}+5 q^{35}+4 q^{37}+4 q^{39} & 57 \\ 
&  +3 q^{41}+3 q^{43}+2 q^{45}+2 q^{47}+q^{49}+q^{51} & \\
      \hline
    \end{array}
    \ee
  \end{small}
  \caption{$q$-weighted numbers of lattice paths under the $y=\frac{3}{3p+1}x$ line. For $q=1$ (right column) these numbers reduce to first nontrivial coefficients given in table \protect\ref{tab-3p2}.}   \label{tab-33p2-qpaths}
\end{table}


\subsection{$(3,3p+1)$ torus knots}

Quiver matrices for $(3,3p+1)$ torus knot have an analogous structure to those in the previous section, and take form
\ea {
  C^{(3,3p+1)} = \begin{bmatrix}
    C^{(2, 6p+1;-2)} & B_p \\
    B_p & C^{(3, 3p-2;+5)}
  \end{bmatrix}
}
where again $C^{(2, 6p+1)}$ denotes a quiver matrix for $(2, 6p+1)$ torus knot, and $C^{(3, 3p-2;+4)}$ is a quiver matrix of $(3, 3p-2)$ torus knot with each entry increased by $+4$ (i.e. with additional framing $+4$). The off-diagonal rectangular block $B$ is organized in the following way 
\ea {
  B_p = \begin{bmatrix}
    B_{p,p} & B_{p,p-1} & \dots & B_{p,1} 
    \end{bmatrix},
}
where each $B_{p,k}$ (for $1 \leq k \leq p$) is a matrix of the size
\ea {
  (3p+1) \times (3k-2).
}
The block $B_{p,k}$ is now composed of three groups of rows
\ea {
  B_{p,k} = \begin{bmatrix}
    X_{p,k} \\
    Y_{p,k} \\
    Z_{p,k}
  \end{bmatrix}
}
with sizes
\ea {
  &X_{p,k}: \quad (p-k+1) \times (3k-2), \\
  &Y_{p,k}: \quad (3k-1) \times (3k-2), \\
  &Z_{p,k}: \quad (2p-2k+1) \times (3k-2).
}
Matrices $X_{p,k}$ and $Y_{p,k}$ have the same structure as for the $(3,3p+2)$ series, however their overall shift is different. Matrix $Z_{p,k}$ consists of rows of constant values, and consecutive rows differ by $1$. Explicitly
\bea
  X_{p,k} &=& \begin{bmatrix}
    6p+3k+1 & 6p+3k-1 & 6p+3k-3 & \cdots & 6k-3k+7 \\
    6p+3k & 6p+3k-2 & 6p+3k-4 & \cdots & 6k-3k+6 \\
    6p+3k-1 & 6p+3k-3 & 6p+3k-5 & \cdots & 6p-3k+5 \\
    \vdots & \vdots & \vdots & \ddots & \vdots \\
    5p+4k+1 & 5p+4k-1 & 5p+4k-3 & \cdots & 5p-2k+7
  \end{bmatrix},
  \\\nonumber\\\nonumber\\
  Y_{p,k} &=& \begin{bmatrix}
    5p+4k & 5p+4k-2  & 5p+4k-4 & \cdots &  5p-2k+6 \\
    5p+4k-1 & 5p+4k-2 & 5p+4k-4 & \cdots & 5p-2k+6 \\
    5p+4k-3 & 5p-4k-3 & 5p+4k-4 & \cdots & 5p-2k+6\\
    \vdots& & \vdots & \vdots & \ddots & \vdots \\
    5p-2k + 54 & 5p-2k+5 & 5p-2k+5 & \cdots & 5p-2k+5
  \end{bmatrix},
  \\\nonumber\\\nonumber\\
  Z_{p,k} &=& \begin{bmatrix}
    5p-2k+4 & 5p-2k+4 & 5p-2k+4 & \cdots & 5p-2k+4 \\
    5p-2k+3 & 5p-2k+3 & 5p-2k+3 & \cdots & 5p-2k+3 \\
    5p-2k+2 & 5p-2k+2 & 5p-2k+2 & \cdots & 5p-2k+2 \\
    \vdots & \vdots & \vdots & \ddots & \vdots \\
    3p+4 & 3p+4 & 3p+4 & \cdots & 3p+4
  \end{bmatrix}.
\eea

Let us also consider first two examples. The value $p=1$ corresponds to $(3,4)$ torus knot. In this case $B_{1,1}$ is built out of $X_{1,1}$, $Y_{1,1}$ and $Z_{1,1}$, and takes form
\ea {
  B_{1,1} = \begin{bmatrix}
 10 \\
 9 \\
 8 \\
 7 \\
  \end{bmatrix}.
}
The full quiver matrix, with the block structure highlighted, reads
\ea {
  C^{(3,4)} =  \begin{bmatrix}
    C^{(2,7;-2)} & B_1 \\
    B_1 & C^{(3,1;+5)}
  \end{bmatrix} =
  \begin{bmatrix}
 13 & 11 & 9 & 7 & 10 \\
 11 & 11 & 9 & 7 & 9 \\
 9 & 9 & 9 & 7 & 8 \\
 7 & 7 & 7 & 7 & 7 \\
 10 & 9 & 8 & 7 & 9 \\
\end{bmatrix},
}
and after reordering columns and rows is equal to quiver presented in~\eqref{quiver34}.
The second example, with $p=2$, corresponds to $(3,7)$ torus knot. In this case we find two matrices 
\ea {
  B_{2,2} = \begin{bmatrix}
 19 & 17 & 15 & 13 \\
 18 & 16 & 14 & 12 \\
 17 & 16 & 14 & 12 \\
 15 & 15 & 14 & 12 \\
 13 & 13 & 13 & 12 \\
 11 & 11 & 11 & 11 \\
 10 & 10 & 10 & 10 \\
  \end{bmatrix}   \qquad \qquad
  B_{2,1} = \begin{bmatrix}
 16 \\
 15 \\
 14 \\
 13 \\
 12 \\
 11 \\
 10 \\
    \end{bmatrix}
}
and the full quiver matrix takes form
\ea {
  C^{(3,7)} = \begin{bmatrix}
    C^{(2,13;-2)} & B_1 \\
    B_1 & C^{(3,4;+5)}
  \end{bmatrix} = 
\begin{bmatrix}
 22 & 20 & 18 & 16 & 14 & 12 & 10 & 19 & 17 & 15 & 13 & 16 \\
 20 & 20 & 18 & 16 & 14 & 12 & 10 & 18 & 16 & 14 & 12 & 15 \\
 18 & 18 & 18 & 16 & 14 & 12 & 10 & 17 & 16 & 14 & 12 & 14 \\
 16 & 16 & 16 & 16 & 14 & 12 & 10 & 15 & 15 & 14 & 12 & 13 \\
 14 & 14 & 14 & 14 & 14 & 12 & 10 & 13 & 13 & 13 & 12 & 12 \\
 12 & 12 & 12 & 12 & 12 & 12 & 10 & 11 & 11 & 11 & 11 & 11 \\
 10 & 10 & 10 & 10 & 10 & 10 & 10 & 10 & 10 & 10 & 10 & 10 \\
 19 & 18 & 17 & 15 & 13 & 11 & 10 & 18 & 16 & 14 & 12 & 15 \\
 17 & 16 & 16 & 15 & 13 & 11 & 10 & 16 & 16 & 14 & 12 & 14 \\
 15 & 14 & 14 & 14 & 13 & 11 & 10 & 14 & 14 & 14 & 12 & 13 \\
 13 & 12 & 12 & 12 & 12 & 11 & 10 & 12 & 12 & 12 & 12 & 12 \\
 16 & 15 & 14 & 13 & 12 & 11 & 10 & 15 & 14 & 13 & 12 & 14 \\
\end{bmatrix}
}
We compute again the classical limit of the generating series~\eqref{definv-class}. In the table~\ref{tab-3p1} we present results for several $(3,3p+1)$ torus knots. These numbers agree with numbers of lattice paths given by the Bizley formula~\eqref{y-Bizley}. 

\begin{table}[h!]
  \centering
\begin{tabular}{|c | l|}
\hline
  \textrm{\bf Knot} & $\qquad b_n$ (numbers of lattice paths) \\ \hline
  (3,4) & 1, 5, 227, 15090, 1182187, 101527596, 9247179818,  \ldots \\
  (3,7) & 1, 12, 2010, 500449, 147412519, 47674321878, 16364395381824,  \ldots \\
  (3,10) & 1, 22, 9097, 5630306, 4129734800, 3328003203564, 2847460237999311,  \ldots \\
  (3,13) & 1, 35, 28931, 35938015, 52957121322, 85769505414732, \dots\\
  (3,16) & 1, 51, 73950, 161559908, 418968975977, \dots \\
\hline
\end{tabular}
\caption{Number of lattice paths under the $y=\frac{3}{3p+1}x$ line, encoded in the classical generating function of $(3,3p+1)$ torus knot.}  \label{tab-3p1}
\end{table}


\section{Full HOMFLY-PT polynomials for the unknot and Schr{\"o}der paths}   \label{sec-schroeder}

So far, in the correspondence with path counting, we considered extremal HOMFLY-PT polynomials. They depend only on one variable $q$, whose powers measure the area in the path interpretation, see fig. \ref{14path}. It is then natural to ask whether the full HOMFLY-PT polynomials also have some path counting interpretation, and if so, what is the interpretation of the variable $a$ in this case. In this section we present a teaser of such an analysis, by discussing the unknot invariants. Note that some other relation between Schr{\"o}der paths and superpolynomials for torus knots was also found in \cite{DMMSS}, and related models are considered in \cite{ORSG} -- it would be interesting to understand if there is some relation between those works and our results.

Recall that the full colored HOMFLY-PT polynomials of the unknot in the trivial framing take form
\be
P_r(a,q) = a^{-r} q^{r} \frac{(a^2; q^2)_r}{(q^2;q^2)}.    \label{Pr-full-unknot}
\ee
Let us consider these invariants in framing $f=1$, which should then correspond to counting paths under the diagonal line $y=x$. In the knots-quivers correspondence, the corresponding quiver was found in \cite{Kucharski:2017ogk} and in $f=1$ framing it takes form
\be
C = \begin{bmatrix}
  2 & 1 \\
  1 & 1
\end{bmatrix}   \label{C-unknot}
\ee
so that (\ref{Pr-full-unknot}) can be obtained from (\ref{P-C}) with the following identification of the variables 
\be
x_1 = - a q^{-1} x,\qquad \quad x_2 = a^{-1} x  \label{unknot_x2}.
\ee
The corresponding generalized A-polynomial for the unknot in framing $f=1$ reads
\be
A(x,y,a) = 1 - y - a^{-1} xy + a x y^2, \label{A-unknot}
\ee
and written in terms of $q=1$ limit of variables (\ref{unknot_x2}) it reads
\be
A(x_1,x_2,a) = 1 - y - x_2y - x_1 y^2. \label{A-unknot-bis}
\ee

\begin{figure}[h]
  \centering
  \includegraphics[scale=0.3]{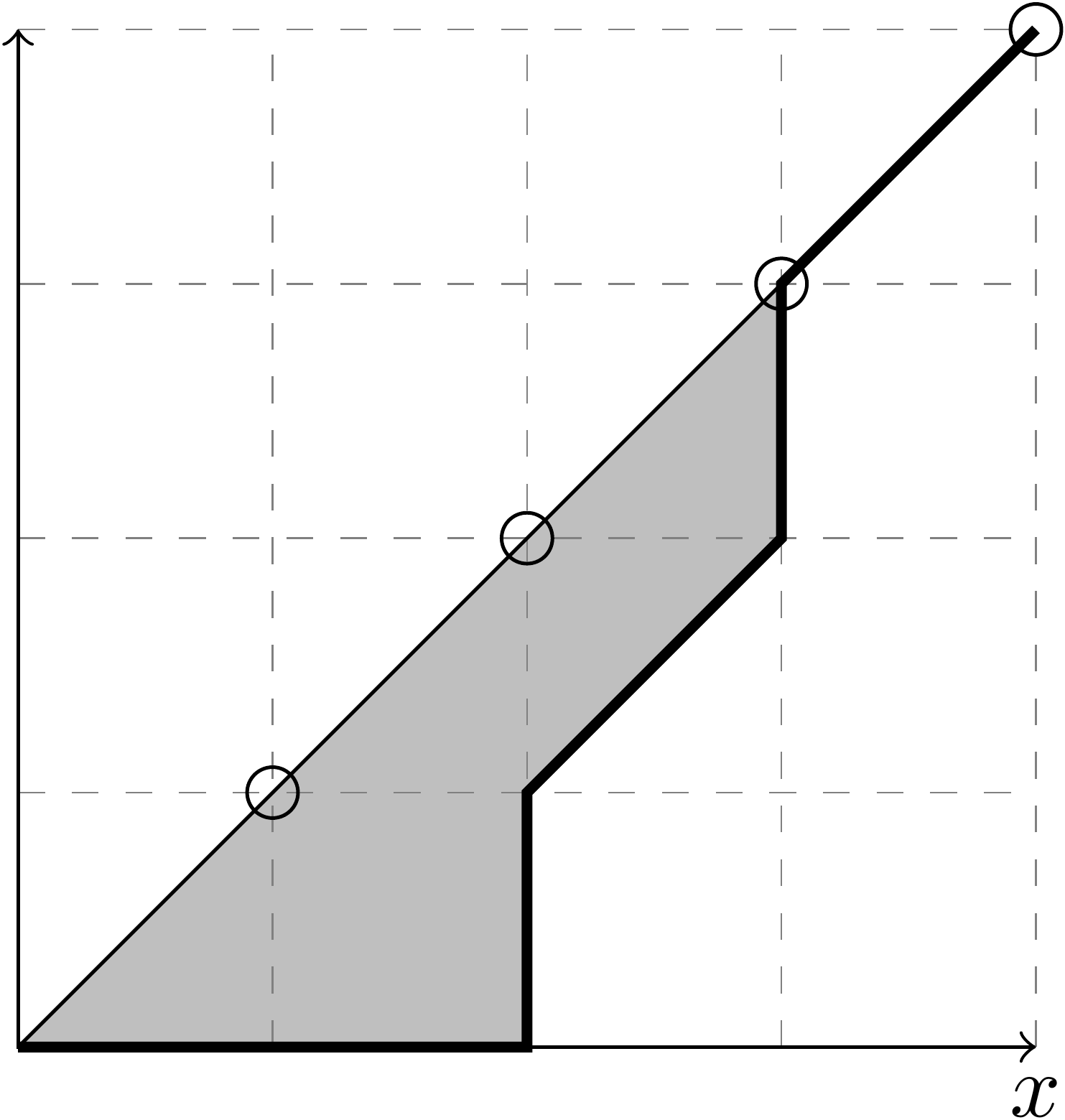}
  \caption{An example of a Schr{\"o}der path of length $6$.}  \label{schroeder}
\end{figure}

For $C$ given in (\ref{C-unknot}), an analogous ratio of quiver generating functions as in (\ref{qPaths-quiver}), with analogous rescaling of generating parameters $x_i\mapsto x_iq^{-1}$, but without setting $x_i$ equal to each other, takes form
\be
y(x_1, x_2,q) = \frac{P_C(x_1 q, x_2 q)}{P_C(x_1 q^{-1}, x_2 q^{-1})} = 1 + q x_1 + x_2 + (q^2 + q^4)x_1^2 + (2q + q^3)x_1 x_2 + x_2^2 + \ldots  \label{yx1x2-schroder}
\ee
Amusingly, this result is related to the $q$-weighted counting of so-called Schr{\"o}der paths. Recall that these are paths made of the two usual steps that we discussed so far, and an additional diagonal step. In the generating function (\ref{yx1x2-schroder}) powers of $x_1$ count the number of steps to the right (of direction $(1,0)$), powers of $x_2$ count number of diagonal steps (of direction $(1,1)$), and powers of $q$ -- as before -- compute the area between the path and the $y=x$ line.  An example of a Schr{\"o}der path is shown in fig.~\ref{schroeder}. To make a direct relation to variables of HOMFLY-PT polynomials, we can rescale $x_1$ and $x_2$ as follows
\be
x_1 = x, \qquad\quad x_2 = ax.
\ee
In such variables (\ref{yx1x2-schroder}) takes form
\be
y(x,a,q) = 1 + (q +a)x + \big( q^2 + q^4  + (2q + q^3)a  + a^2\big) x^2 + \ldots  \label{yx1x2-schroder-2}
\ee
with the length of a path measured by the power of $x$ and the number of diagonal steps measured by the power of $a$.

\begin{figure}[h]
  \includegraphics[scale=0.3]{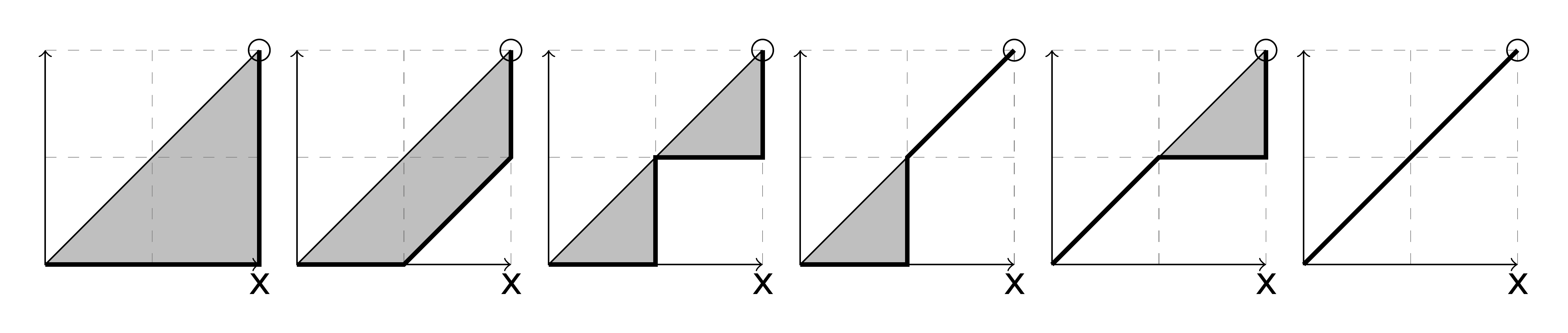}
  \caption{All $6$ Schr{\"o}der paths represented by quadratic terms $(q^2 + q^4)x_1^2 + (2q + q^3)x_1 x_2 + x_2^2$ of the generating function (\protect\ref{yx1x2-schroder}).}  \label{schroeder2}
\end{figure}

There are several interesting limits of (\ref{yx1x2-schroder}). In the homogenous classical limit we get
\be
y(x, x, 1) = 1 + 2x + 6x^2 + 22x^3 + 90 x^4 + 394 x^5 + \ldots
\ee
and the coefficients of this series simply count all Schr{\"o}der paths of a given height. For example, 6 paths of height 2 are shown in fig. \ref{schroeder2} (more generally, the area measured by powers of $q$ in (\ref{yx1x2-schroder}) is shown in grey). On the other hand, setting $x_2=0$ we obtain the generating function of $q$-Catalan numbers, which reproduce the result from section \ref{ssec-unknot}
\be
y(x_1,0, q)  = 1 + q x_1 +  (q^2 + q^4)x_1^2   + \ldots   
\ee
Finally, for $x_1=0$ we get a geometric series representing only all diagonal paths (with vanishing area)
\be
y(0, x_2,q) =  1  + x_2 +   x_2^2 + \ldots   = \frac{1}{1-x_2}.
\ee
One can easily check, that all above statements hold also for the $f$-framed unknot that corresponds to Schr{\"o}der under the line $y=fx$; this provides a generalization of results mentioned in section \ref{ssec-unknot} to the $a$-deformed case.

As in the previous cases, the A-polynomial (\ref{A-unknot}) can be reproduced, from the path counting perspective, from the Duchon grammar, which now consists of three letters. It reads
\be
A(x,y_P) = 1 + (x-1)y_P + x y_P^2, \label{A-unknot-bisbis}
\ee
where the term $xy_P$ is due to the horizontal step and $x y_P^2$ due to the two ascending and descending steps. Up to powers of $a$, this result indeed agrees with ~\eqref{A-unknot}. In this interpretation the role of the variable $a$ is to distinguish paths of the same length but with different number of horizontal steps. Equivalently, (\ref{A-unknot-bisbis}) is directly identified with (\ref{A-unknot-bis}) upon the identification $x=-x_1=-x_2$ and $y=y_P$.

In summary, at least in the case of the unknot, introducing the variable $a$ of HOMFLY-PT polynomials corresponds to adding an additional diagonal step in the path counting interpretation. We postpone the generalization of this picture to other torus knots to future work.


\acknowledgments{We thank Adam Doliwa, Eugene Gorsky, Sergei Gukov, Piotr Kucharski, and Markus Reineke for their interest in this work, useful comments and enlightening discussions. Parts of this work have been done while M.S. and P.S. were visiting Max-Planck Institute for Mathematics (Bonn, Germany), American Institute for Mathematics (San Jose, USA), and Isaac Newton Institute for Mathematical Sciences (Cambridge, UK). This work is supported by the ERC Starting Grant no. 335739 \emph{``Quantum fields and knot homologies''} funded by the European Research Council under the European Union's Seventh Framework Programme, and the Foundation for Polish Science. The work of M.S. was also partially supported by the Portuguese Funda\c c\~ao para a Ci\^encia e a Tecnologia (FCT) through the FCT Investigador Grant, and also by the Ministry of Education, Science, and Technological Development of the Republic of Serbia, project no. 174012. M.P. acknowledges the support from the National Science Centre through the FUGA grant 2015/16/S/ST2/00448}.



\newpage

\bibliographystyle{JHEP}
\bibliography{abmodel}

\end{document}